\documentclass[12pt,epsf]{article}
\usepackage{graphicx,epsf}
\usepackage{lscape}
\usepackage{citesort}

\begin{document}

\def\ds{\displaystyle}
\def\beq{\begin{equation}}
\def\eeq{\end{equation}}
\def\bea{\begin{eqnarray}}
\def\eea{\end{eqnarray}}
\def\beeq{\begin{eqnarray}}
\def\eeeq{\end{eqnarray}}
\def\ve{\vert}
\def\vel{\left|}
\def\ver{\right|}
\def\nnb{\nonumber}
\def\ga{\left(}
\def\dr{\right)}
\def\aga{\left\{}
\def\adr{\right\}}
\def\lla{\left<}
\def\rra{\right>}
\def\rar{\rightarrow}
\def\nnb{\nonumber}
\def\la{\langle}
\def\ra{\rangle}
\def\ba{\begin{array}}
\def\ea{\end{array}}
\def\tr{\mbox{Tr}}
\def\ssp{{\Sigma^{*+}}}
\def\sso{{\Sigma^{*0}}}
\def\ssm{{\Sigma^{*-}}}
\def\xis0{{\Xi^{*0}}}
\def\xism{{\Xi^{*-}}}
\def\qs{\la \bar s s \ra}
\def\qu{\la \bar u u \ra}
\def\qd{\la \bar d d \ra}
\def\qq{\la \bar q q \ra}
\def\gGgG{\la g^2 G^2 \ra}
\def\q{\gamma_5 \not\!q}
\def\x{\gamma_5 \not\!x}
\def\g5{\gamma_5}
\def\sb{S_Q^{cf}}
\def\sd{S_d^{be}}
\def\su{S_u^{ad}}
\def\ss{S_s^{??}}
\def\sbp{{S}_Q^{'cf}}
\def\sdp{{S}_d^{'be}}
\def\sup{{S}_u^{'ad}}
\def\ssp{{S}_s^{'??}}
\def\sig{\sigma_{\mu \nu} \gamma_5 p^\mu q^\nu}
\def\fo{f_0(\frac{s_0}{M^2})}
\def\ffi{f_1(\frac{s_0}{M^2})}
\def\fii{f_2(\frac{s_0}{M^2})}
\def\O{{\cal O}}
\def\sl{{\Sigma^0 \Lambda}}
\def\es{\!\!\! &=& \!\!\!}
\def\ap{\!\!\! &\approx& \!\!\!}
\def\ar{&+& \!\!\!}
\def\ek{&-& \!\!\!}
\def\kek{\!\!\!&-& \!\!\!}
\def\cp{&\times& \!\!\!}
\def\se{\!\!\! &\simeq& \!\!\!}
\def\eqv{&\equiv& \!\!\!}
\def\kpm{&\pm& \!\!\!}
\def\kmp{&\mp& \!\!\!}


\def\simlt{\stackrel{<}{{}_\sim}}
\def\simgt{\stackrel{>}{{}_\sim}}


\title{ {\Large {\bf Semileptonic $B _{c}^{-}\rightarrow D^{*0}\ell\nu$ transition in three--point QCD sum rules and HQET with gluon
condensate corrections } } }

\author{\vspace{1cm}\\
{\small N. Ghahramany$^1$ \thanks {e-mail:ghahramany @ susc.ac.ir
}~\,},
{\small R. Khosravi$^1$ \thanks {e-mail: khosravi.reza @ gmail.com}~\,}, {\small K. Azizi $^2$\thanks {e-mail: e146342 @ metu.edu.tr}~\,}\\
 {\small $^1$ Physics Department , Shiraz University, Shiraz 71454,
Iran}\\
{\small  $^2$ Department of Physics, Middle East Technical
University, 06531 Ankara, Turkey}\\}
\date{}

\begin{titlepage}
\maketitle \thispagestyle{empty}

\begin{abstract}
Taking into account the gluon condensate contributions, the form
factors of the semileptonic $B_c^- \rightarrow D^{*0}\ell\nu$
transition with $l=\tau, e$  are calculated in the framework of the
three point QCD sum rules. The heavy quark effective theory limit of
the form factors are also computed. The relevant total decay width
as well as the branching ratio  are  evaluated and compared with the
predictions of the other non-perturbative approaches.
\end{abstract}


\end{titlepage}

\section{Introduction}

In 1998, the CDF collaboration reported  the first experimental
observation of the $B_c$ meson\cite{1}. The heavy meson $B_c$ with
$\bar{b}c$ quark structure is made of two heavy quarks with
different charge and flavors. It is located between two heavy meson
families called charmonium $\bar{c}c$ and bottomonium $\bar{b}b$, so
this meson is similar to the the charmonium and bottomonium in the
spectroscopy. The  predictions for the mass spectra of the 
$\bar{b}c$~ levels were obtained in the potential models (PM) and
lattice simulations \cite{2,3,4,5,6}. In contrast to the charmonium
and bottomonium, the  $B_c$ decays only via weak interaction and holds
long life time. For this reason the $B_c$ transitions are very
interesting tool to  calculate more precise values for the
Cabibbo-Kabayashi-Maskawa(CKM) matrix elements and  study the CP and
T violations that occur in weak interactions. It is predicted that
the LHC experiments may give interesting informations about this
meson that could be used as a basis for future
investigations\cite{7,8}.

Some decay modes of this meson have been studied by
different methods. The $B_c\rightarrow D_{s,d} \ell^{+}
\ell^{-}/ \nu \bar{\nu}$, $B_c\rightarrow D^{*}_{s,d} \ell^{+}
\ell^{-}$, $B_c\rightarrow X (D^*,D_s^*,D_1,D_{s1})\nu \bar{\nu}$ transitions have been discussed
via three point QCD sum rules (3PSR) in \cite{9,10,11}, the
$B_c\rightarrow j/\psi \ell \nu $ has been analyzed by means of the  3PSR and non-relativistic QCD (NRQCD) \cite{40} and the 
$B_c\rightarrow \ell \bar{\nu}\gamma $, $ B_c\rightarrow
\rho^{+}\gamma$,
 $B_{c}\rightarrow K^{\ast+}\gamma$, $B_c\rightarrow B_u^{*}
\ell^{+} \ell^{-}$ channels  have been investigated in the framework of the  light-cone QCD sum rules
\cite{13,14,15,16}. A large set of the exclusive non-leptonic and
semileptonic decays of the $B_c$ meson have been studied within the
potential model (PM) (see \cite{17,18,19,20,21,22,23,24,25,26}),
and also operator product expansion in inverse powers of the heavy
quark masses \cite{27}. In this work, considering the gluon
corrections to the  relevant form factors,  the  $B_c^-\rightarrow
D^{*0} \ell \nu$ mode is investigated in the framework of  the
three-point QCD sum rules (3PSR) and also in the heavy quark
effective theory (HQET). This decay mode has been discussed in
different methods (for instance see\cite{12,21,24,28,30,32}). This
transition has also been investigated in the QCD sum rule
approach, for example in \cite{12,29} but without considering the
gluon corrections. In \cite{12}, the coulomb like corrections were
considered in the calculations to decrease the uncertainties. The
main points in the present work  are the calculation of the gluon
corrections and check whether their contributions guarantee the
convergence of the sum rules for the form factors or not and also
the comparison between the form factors and their HQET limit. For
this aim, we plot the dependence of both form factors and their
HQET limit on the transferred momentum square ($q^2$) and compare
them at high and low $q^2$ values.

This paper includes five sections. The calculation of the sum rules
for the relevant form factors are presented in Section 2. In the sum
rules expressions for the form factors, the light quark condensate
do not have any contributions since applying the double Borel
transformation with respect to the momentum of the initial and final
states kills their contributions. Therefore, as a first correction
on the non-perturbative part of the correlation function, the two
gluon condensate contributions  are taken into account, so in
Section3, the gluon condensate contributions in the Borel transform
scheme is presented. Section 4 is devoted to the explanation of the
heavy quark effective theory. In this section HQET limit of the form
factors are derived. Section 5 depicts our numerical analysis
of the form factors and their comparison with the HQET limit.
  This section also contains the calculation of the total decay width as well as the branching ratio of the $B_c^-\rightarrow
D^{*0} \ell \nu ~(\ell=e,\tau)$ via 3PSR and HQET and their
comparison  with the predictions of other approaches.

\section{Sum rules for $B_{c}^-\rightarrow D^{*0}\ell\nu $transition form factors}

At quark level, the tree level $b\rightarrow u$ transition is
responsible for $B_c^-\rightarrow  D^{*0}$ decay mode. The
Hamiltonian of this decay is written as:
\begin{equation}  \label{eq1}
\mathcal{H}_{q}=\frac{G_{F}}{\sqrt{2}} V_{ub}~\overline{\nu}
~\gamma_{\mu}(1-\gamma_{5})l~\overline{u} ~\gamma_{\mu}(1-\gamma_{5}) b, \\
\end{equation}
where, $G_{F}$ is the Fermi constant and $V_{ub}$ is the CKM matrix
element. The decay amplitude for $B_c^-\rightarrow D^{*0} \ell \nu$
is obtained by sandwiching the Eq. (\ref{eq1}) between the initial
and final meson states
\begin{equation}  \label{eq2}
\mathcal{M}=\frac{G_{F}}{\sqrt{2}} V_{ub}~\overline{\nu} ~\gamma_{\mu}(1-%
\gamma_{5})l\langle D^{*}(p^{\prime},\varepsilon)\mid~\overline{u} ~\gamma_{\mu}(1-%
\gamma_{5}) b\mid B_{c}(p)\rangle.
\end{equation}

Our main task is to calculate the matrix element appearing in the
Eq.(\ref{eq2}). Both axial and vector parts of the transition current are involved in
this matrix element. Using the Lorentz invariance and parity
conservation, it can be parameterized in terms of some form factors as:
\begin{equation}  \label{eq3}
<D^*(p^{\prime},\varepsilon)\mid\overline{u}\gamma_{\mu}
b\mid B_c(p)>=-f^{'}_{V}(q^2)\varepsilon_{\mu\nu\alpha\beta}
\varepsilon^{\ast\nu}p^\alpha {p^{\prime}}^\beta,
\end{equation}
\begin{eqnarray}  \label{eq4}
< D^*(p^{\prime},\varepsilon)\mid\overline{u}\gamma_{\mu}\gamma_{5} b\mid
B_c(p)> &=&-i\left[f^{'}_{0}(q^2)\varepsilon_{\mu}^{\ast} \right.
\nonumber \\
+ f^{'}_{1}(q^2)(\varepsilon^{\ast}p)P_{\mu} &+&
\left. f^{'}_{2}(q^2)(\varepsilon^{\ast}p)q_{\mu}%
\right],
\end{eqnarray}

where:
\begin{eqnarray}  \label{eq5}
f_{V}^{\prime}(q^2)&=&\frac{2f_{V}(q^2)}{(m_{B_c}+m_{D^*})}%
~,~~~~~~~~~~~~f_{0}^{\prime}(q^2)=f_{0}(q^2)(m_{B_c} +m_{D^*}),
\nonumber
\\
f_{1}^{\prime}(q^2)&=&-\frac{f_{1}(q^2)}{(m_{B_c}+m_{D^*})}~,~~~~~~~~~~~~
f_{2}^{\prime}(q^2)=-\frac{f_{2}(q^2)}{(m_{B_c}+m_{D^*})},
\end{eqnarray}
 and the $f_{V}(q^2)$, $f_{0}(q^2)$, $f_{1}(q^2)$ and $f_{2}(q^2)$ are the
transition form factors,  $P_{\mu}=(p+p^{\prime})_{\mu}$, $%
q_{\mu}=(p-p^{\prime})_{\mu}$ and $\varepsilon$ is the
four--polarization vector of the $D^*$ meson.

To find the above form factors via sum rules, we use  the following three-point correlation function:
\begin{eqnarray}\label{eq6}
\Pi_{\mu\nu} (p,p^\prime,q) = i^2 \int d^4 x d^4 y e^{+i p^\prime
x - i p y} \left< 0 \left| \mbox{\rm T} \left\{ J_{D^{*} \nu}(x)
J_\mu (0)  J^{\dagger}_{B_{c}}(y)\right\} \right| 0\right>~,
\end{eqnarray}
where $J_{D^{*}\nu}(x)=\overline{c}\gamma_{\nu}u$,
 $J_{B_{c}}(y)=\overline{c}\gamma_{5}b$,
 $J_{\mu}^{V}=\overline{u}\gamma_{\mu}b$ and
 $J_{\mu}^{A}=\overline{u}\gamma_{\mu}\gamma_{5}b$ are the
interpolating currents of the $ D^* $, $B_{c} $, vector and axial
vector parts of the transition current, respectively.

The following relation hold for the Lorentz structures of the
selected correlation functions:
\begin{eqnarray}\label{eq7}
\Pi_{\mu\nu}^{(V-A)} = \epsilon_{\mu\nu\alpha\beta} \,p^\alpha
p^{\prime\beta} \Pi_{f_V}+ \Pi_{f_0} g_{\mu\nu} + \Pi_{f_1} {\cal
P}_\mu p_\nu + \Pi_{f_2} q_\mu p_\nu + ...
\end{eqnarray}

By inserting two complete sets of the intermediate states with the same
quantum number as the currents $J_{D^*}$ and $J_{B_c}$, we can
calculate the phenomenological part of the correlators  given in the 
Eq. (\ref{eq6}) as follow:
\begin{eqnarray} \label{eq8}
&&\Pi _{\mu\nu}^{V-A}(p^2,p'^2,q^2)=
\nonumber \\
&& \frac{<0\mid J_{D^* \nu} \mid D^*(p',\varepsilon)><
D^*(p',\varepsilon)\mid J_{\mu}^{V-A}\mid B_{c}(p)><B_{c}(p)\mid
J_{Bc}\mid0>}{(p'^2-m_{D^*}^2)(p^2-m_{Bc}^2)}+\cdots
\nonumber \\
\end{eqnarray}

where... denotes the contributions coming  from the higher states and
continuum. In the  Eq. (\ref{eq8}), the vacuum to the initial and final meson
state matrix elements are defined  as:
\begin{eqnarray}  \label{eq9}
<0\mid J_{D^{*}}^{\nu}\mid D^{*}(p^{\prime
})>=f_{D^*}m_{D^*}\varepsilon ^{\nu }~,~~
<0\mid J_{B_c}\mid B_{c}(p)>=i\frac{f_{B_c%
}m_{B_c}^{2}}{m_{b}+m_{c}},
\end{eqnarray}

where $f_{D^*}$ and $f_{B_c}$ are the leptonic decay constants of
the $D^{*}$ and $B_c$ mesons, respectively. Using the Eqs. (\ref{eq3}),
(\ref{eq4}) and (\ref{eq9}) in the Eq. (\ref{eq8}) and performing
summation over the polarization of the  $D^{*}$ meson, we get the
following result for the physical part:
\begin{eqnarray}\label{eq10}
\Pi_{\mu\nu}^{V}(p^2,p'^2,q^2)&=&-\frac{f_{B_c}m_{B_c}^2}{(m_{b}+m_{c})}\frac{f_{D^*}m_{D^*}}
{(p'^2-m_{D^*}^2)(p^2-m_{B_c}^2)} \times
[f_{0}'g_{\mu\nu}+f_{1}'P_{\mu}p_{\nu} \nonumber
\\ &+&f_{2}'q_{\mu}p_{\nu}]+
\mbox{excited states.}\nonumber
\\\Pi_{\mu\nu}^{A}(p^2,p'^2,q^2)&=&
-i\varepsilon_{\mu\nu\alpha\beta}p'^{\alpha}p^{\beta}\frac{f_{B_c}m_{B_c}^2}{(m_{b}+m_{c})}\frac{f_{D^*}m_{D^*}}
{(p'^2-m_{D^*}^2)(p^2-m_{B_c}^2)}f_{V}' +
\nonumber \\
&&\mbox{excited states.}
\end{eqnarray}
The coefficients of the Lorentz structures
$\epsilon_{\mu\nu\alpha\beta}p^{\alpha}p^{'\beta}$, $ig_{\mu\nu}$,
$P_{\mu}p_{\nu}$ and $q_{\mu}p_{\nu}$ in the correlation functions
$\Pi_{\mu\nu}^{V}$and $\Pi_{\mu\nu}^{A}$ will be chosen  in
determination of the form factors
$f_{V}(q^2)$,~$f_{0}(q^2)$,~$f_{1}(q^2)$ and $f_{2}(q^2)$,
respectively.

The QCD side of the correlation function is calculated with the help
of  the operator product expansion (OPE) in the deep Euclidean
region where $p^2\ll (m_b+m_c)^2$, $p'^2\ll m_c^2$. In the 
Eq. (\ref{eq6}), using the expansion of the time ordered products of
currents  in terms
of a series of local operators with increasing dimension, we will
have \cite{15}:
\begin{eqnarray}\label{eq11}
-\int d^4x d^4y e^{-i(p x - p^\prime y)} T \Big\{J_{D^{*}\nu}
J_\mu J_{B_c} \Big\} \es (C_0)_{\nu\mu} I + (C_3)_{\nu\mu} \bar{q}
q + (C_4)_{\nu\mu} G_{\alpha\beta}
G^{\alpha\beta} \nnb \\
\ar (C_5)_{\nu\mu} \bar{q} \sigma_{\alpha\beta} G^{\alpha\beta} q
+ (C_6)_{\nu\mu} \bar{q} \Gamma q \bar{q} \Gamma^\prime q~,
\end{eqnarray}

where $(C_i)_{\mu\nu}$ are the Wilson coefficients,
$G_{\alpha\beta}$ is the gluon field strength tensor, $I$ is the
unit operator, $\Gamma $ and $\Gamma^{'}$ are the matrices
appearing in the calculations. Taking into account the vacuum
expectation value of the OPE, the expansion of the correlation function
in terms of the local operators is written as follow:
\begin{eqnarray}\label{eq12}
\Pi_{\nu\mu} (p_1^2,p_2^2,q^2) \es
(C_0)_{\nu\mu} + (C_3)_{\nu\mu} \la \bar{q} q \ra + (C_4)_{\nu\mu}
\la G^2 \ra + (C_5)_{\nu\mu} \la \bar{q}
\sigma_{\alpha\beta} G^{\alpha\beta} q \ra \nnb \\
\ar (C_6)_{\nu\mu} \la \bar{q} \Gamma q \bar{q} \Gamma^\prime q
\ra~.
\end{eqnarray}

The heavy quark condensate contributions are
suppressed by inverse of the heavy quark mass and can be safely
omitted. The light $u$ quark condensate contributions  are zero after
applying the  double Borel transformation with respect to the   momentum of the initial and final states because only one of them appears in
the denominator (see in Fig. 1(b,c)).

\begin{figure}[th]

\vspace*{4.cm}
\begin{center}
\begin{picture}(160,30)
\centerline{ \epsfxsize=16cm \epsfbox{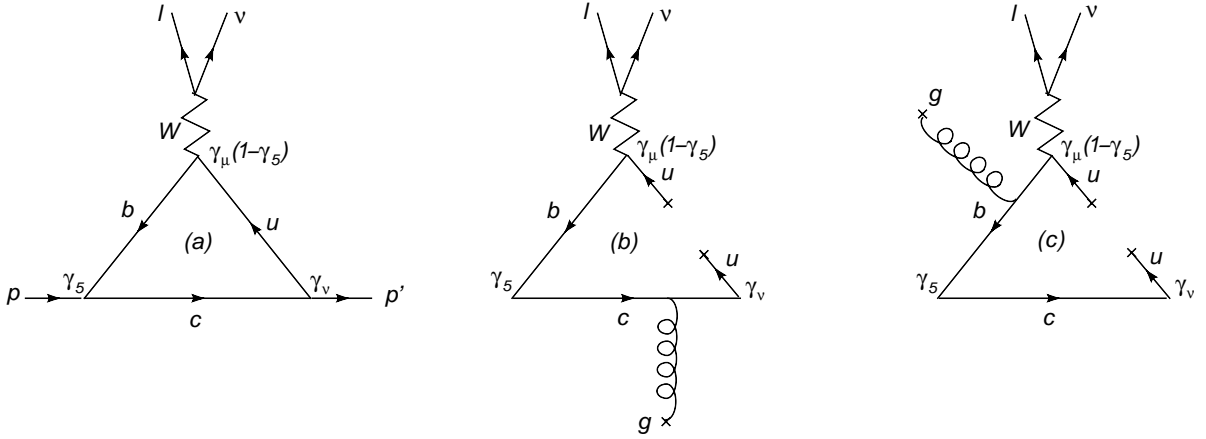}}
\end{picture}
\end{center}

\vspace*{-1cm} \caption{loop diagrams for $B_c \rightarrow
D^{*}\ell\nu$ transitions, bare loop (diagram a) and light quark
condensates with one gluon emission (diagrams b, c)}
\end{figure}
\normalsize

\setlength{\unitlength}{1mm}

As a result,  the
correlation functions receive contribution from the bare-loop,
Fig. 1(a) and gluon condensates, Fig. 2(a-f) i.e.,
\begin{eqnarray}\label{eq13}
\Pi_i(p_1^2,p_2^2,q^2) = \Pi_i^{per}(p_1^2,p_2^2,q^2) + \Pi_i^{\la
G^2 \ra} (p_1^2,p_2^2,q^2) \frac{\alpha_s}{\pi} \la G^2 \ra~.
\end{eqnarray}

Using the double dispersion representation, the bare-loop
contribution is determined:
\begin{eqnarray}\label{eq14}
\Pi_i^{per} = - \frac{1}{(2 \pi)^2} \int \int\frac{\rho_i
(s,s^\prime, q^2)}{(s-p^2) (s^\prime - p^{\prime 2})} ds ds^\prime +
\mbox{\rm subtraction terms}~.
\end{eqnarray}

The following non-equalities give the integration
limits of the Eq. (\ref{eq14}):
\begin{eqnarray}\label{eq15}
-1 \le \frac{2 s s^\prime + (s + s^\prime - q^2 )(m_b^2 - m_c^2 -
s) + 2 s (m_c^2-m^2_{u})}{\lambda^{1/2} (s,s^\prime,q^2)
\lambda^{1/2}(m_b^2,m_c^2,s)} \le +1~,
\end{eqnarray}
where $\lambda(a,b,c)=a^2+b^2+c^2-2ab-2ac-2bc$.

By replacing the propagators with the Dirac-delta
functions(Gutkovsky rule):
\begin{eqnarray}\label{eq16}
\frac{1}{k^2-m^2} \rar -2i\pi \delta(k^2-m^2)~,
\end{eqnarray}

the spectral densities $\rho_i(s,s^{'},q^2)$ are found as:
\begin{eqnarray}  \label{eq17}
\rho _{V} &=&-4\,{N_c}~{I_{0}}(s,s^{\prime },q^2)~\Bigg\{ {m}_{c}+{B_{2}}\,({m%
}_{c}-\,{m_{u})+B_{1}}\,({m}_{c}-{m_{{b}}})\Bigg\} ~,  \nonumber \\
\rho _{0} &=&-2\,N_c~{I_{0}}(s,s^{\prime },q^2)~\Bigg\{4\,{A_{1}}\,({m_{{b}}}-%
{m}_{c})+{\Delta ^{\prime }}\,({m_{{b}}}-{m}_{c})-\Delta
\,({m}_{c}+{m_{u})}
\nonumber \\
&&+2\,{m_{{c}}}^{2}({m_{{c}}-m_{{b}}}-{m_{u}})+{m}_{c}{(u+}2\,{m_{{b}}m_{u})%
\Bigg\}}\,~,  \nonumber \\
\rho _{1} &=&-2\,N_c~{I_{0}}(s,s^{\prime },q^2)\,~\Bigg\{B_{1}(\,{m_{{b}}}-3{m%
}_{c})+2\,{A_{2}}\,({m_{{b}}}-{m}_{c})+2\,{A_{3}}\,({m_{{b}}}-{m}_{c})
\nonumber \\
&&+{B_{2}}\,({m_{u}}-\,{m_{c})}-{m}_{c}\Bigg\}~,  \nonumber \\
\rho _{2} &=&-2\,N_c~{I_{0}}(s,s^{\prime },q^2)\,~\Bigg\{2\,{A_{3}(m}_{c}-\,{%
m_{{b}}})-{B_{1}}\,({m_{{b}}}+{m}_{c})+2\,{A_{2}}\,({m_{{b}}}-{m}_{c})
\nonumber \\
&&+{B_{2}(m}_{c}-{m_{u})}+\,{m}_{c}\Bigg\}~.  \nonumber \\
&&
\end{eqnarray}

where
\begin{eqnarray*}  \label{eq18}
I_{0}(s,s^{\prime},q^2)&=&\frac{1}{4\lambda^{1/2}(s,s^{\prime},q^2)},
\nonumber \\
\lambda(s,s^{\prime},q^2)&=&s^2+{s^{\prime}}^2+q^4-2sq^2-2s^{%
\prime}q^2-2ss^{\prime},  \nonumber \\
B_{1}&=&\frac{1}{\lambda(s,s^{\prime},q^2)}[2s^{\prime}\Delta-\Delta^{%
\prime}u],  \nonumber \\
B_{2}&=&\frac{1}{\lambda(s,s^{\prime},q^2)}[2s\Delta^{\prime}-\Delta
u],
\nonumber \\
A_{1}&=&\frac{1}{2\lambda(s,s^{\prime},q^2)}[{\Delta^{\prime}}%
^{2}s+\Delta^2s^{\prime}-
4m_{u}^2ss^{\prime}-\Delta\Delta^{\prime}u+m_{c}^2u^2],  \nonumber \\
A_{2}&=&\frac{1}{\lambda^{2}(s,s^{\prime},q^2)}[2{\Delta^{\prime}}%
^2ss^{\prime}+6\Delta^2{s^{\prime}}^2 -8m_{c}^2s{s^{\prime}}%
^2-6\Delta\Delta^{\prime}s^{\prime}u  \nonumber \\
&& +{\Delta^{\prime}}^2u^2 +2m_{c}^2s^{\prime}u^2],  \nonumber \\
A_{3}&=&\frac{1}{\lambda^{2}(s,s^{\prime},q^2)}[-3\Delta^2us^{\prime}-3{%
\Delta^{\prime}}^2us
+4m_{c}^2us^{\prime}s+4\Delta\Delta^{\prime}ss^{\prime}
\nonumber \\
&& +2\Delta\Delta^{\prime}u^2 -m_{c}^2u^3].  \nonumber \\
\end{eqnarray*}

The $N_c=3$ is the color factor, $u=s+s^\prime-q^2$,
$\Delta=s+m_c^2-m_b^2$ and $\Delta'=s^\prime +m_c^2-m^2_{u}$.

\section{Gluon condensate contribution}
In this section, the gluon condensate contributions related to the non-perturbative
part of the QCD sum rules  are discussed. The diagrams for contributions of the
gluon condensates are depicted in Fig. 2.

\begin{figure}[th]

\vspace*{4.cm}
\begin{center}
\begin{picture}(160,70)
\centerline{ \epsfxsize=16cm \epsfbox{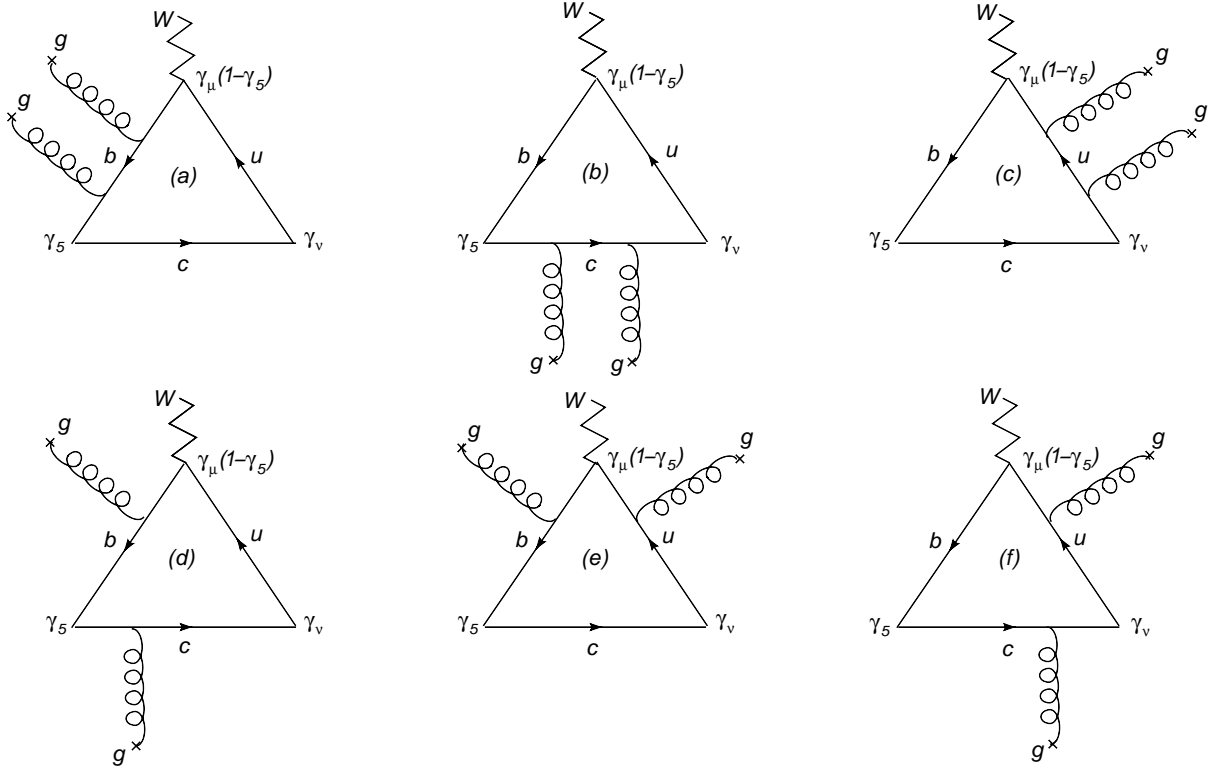}}
\end{picture}
\end{center}

\vspace*{-1cm} \caption{Gluon condensate contributions to $B_c
\rightarrow D^{*}\ell\nu$ transitions  }
\end{figure}
\normalsize

\setlength{\unitlength}{1mm}

To calculate these diagrams, the Fock-Schwinger fixed-point gauge,
$x^\mu G_\mu^a=0$, are used, where $G_\mu^a$ is the gluon field. In
the evaluation of the diagrams in Fig. 2, integrations of the following
type are encountered.
\begin{eqnarray}\label{eq19}
I_{\mu\nu...\tau}(a,b,c) = \int\frac{d^4 k}{(2\pi)^4}
\frac{k_{\mu}k_{\nu}...k_{\tau}}{[k^2 - m_b^2]^a[(p + k)^2 -
m_c^2]^b [(p^{'} + k)^2 - m_u^2]^c}.
\end{eqnarray}

In our case, the following three types of  integrals are appeared:
\begin{eqnarray}  \label{eq20}
I_0(a,b,c) \es \int \frac{d^4k}{(2 \pi)^4} \frac{1}{\left[
k^2-m_b^2 \right]^a \left[ (p+k)^2-m_c^2 \right]^b \left[
(p^\prime+k)^2-m_{u}^2\right]^c}~,
\nnb \\ \nnb \\
I_\mu(a,b,c) \es \int \frac{d^4k}{(2 \pi)^4} \frac{k_\mu}{\left[
k^2-m_b^2 \right]^a \left[ (p+k)^2-m_c^2 \right]^b \left[
(p^\prime+k)^2-m_{u}^2\right]^c}~,
\nnb \\ \nnb \\
I_{\mu\nu}(a,b,c) \es \int \frac{d^4k}{(2 \pi)^4} \frac{k_\mu
k_\nu}{\left[ k^2-m_b^2 \right]^a \left[ (p+k)^2-m_c^2 \right]^b
\left[ (p^\prime+k)^2-m_{u}^2\right]^c}~.
\end{eqnarray}

These integrals can be calculated by continuing to Euclidean
space-time and using Schwinger representation for the Euclidean
propagator
\begin{eqnarray*}  \label{eq21}
\frac{1}{(k^2+m^2)^n} = \frac{1}{\Gamma(n)} \int_0^\infty d\alpha \,
\alpha^{n-1} e^{-\alpha(k^2+m^2)}~,
\end{eqnarray*}

which fits the Borel transformation because:
\begin{eqnarray*}\label{eq22}
{\cal B}_{\hat{p}^2} (M^2) e^{-\alpha p^2} = \delta
(1/M^2-\alpha)~.
\end{eqnarray*}

In order to obtain the Borel transformed form of the integrals in the 
Eq. (\ref{eq20}), the  integration is performed over loop momentum and
over the two parameters used in the exponential representation of the 
propagators:
\begin{eqnarray}\label{eq23}
\hat{I}_{0}(a,b,c)\!\!\! &=&\!\!\!\frac{(-1)^{a+b+c}}{16\pi
^{2}\,\Gamma
(a)\Gamma (b)\Gamma (c)}(M_{1}^{2})^{2-a-b}(M_{2}^{2})^{2-a-c}\,\mathcal{U}%
_{0}(a+b+c-4,1-c-b)~, \nnb \\ \nnb \\
\hat{I}_{\mu }(a,b,c) &=&\hat{I}_{1}(a,b,c)p_{\mu
}+\hat{I}_{2}(a,b,c)p_{\mu
}^{\prime }~, \nnb \\ \nnb \\
\hat{I}_{\mu \nu }(a,b,c) &=&\hat{I}_{6}(a,b,c)g_{\mu \nu }+\hat{I}%
_{3}p_{\mu }p_{\nu }+\hat{I}_{4}p_{\mu }p_{\nu }^{\prime }+\hat{I}%
_{4}p_{\mu }^{\prime }p_{\nu }+\hat{I}_{5}p_{\mu }^{\prime }p_{\nu
}^{\prime }~.
\end{eqnarray}
$\hat{I}$ in the Eq.(\ref{eq23}) stands for the double Borel transformed
form of the Eq.(\ref{eq20}), in the Schwinger representation.

where:
\begin{eqnarray}\label{eq24}
\hat{I}_{1(2)}(a,b,c) \es i \frac{(-1)^{a+b+c+1}}{16
\pi^2\,\Gamma(a) \Gamma(b) \Gamma(c)}
(M_1^2)^{1-a-b+1(2)} (M_2^2)^{4-a-c-1(2)} \, {\cal U}_0(a+b+c-5,1-c-b)~, \nnb \\ \nnb \\
\hat{I}_j(a,b,c) \es i \frac{(-1)^{a+b+c+1}}{16 \pi^2\,\Gamma(a)
\Gamma(b) \Gamma(c)}
(M_1^2)^{-a-b-1+j} (M_2^2)^{7-a-c-j} \, {\cal U}_0(a+b+c-5,1-c-b)~, \nnb \\ \nnb \\
\hat{I}_6(a,b,c) \es i \frac{(-1)^{a+b+c+1}}{32 \pi^2\,\Gamma(a)
\Gamma(b) \Gamma(c)} (M_1^2)^{3-a-b} (M_2^2)^{3-a-c} \, {\cal
U}_0(a+b+c-6,2-c-b)~,
\end{eqnarray}

Here, $j=3,4,5$, $M_1^2$ and $M_2^2$ are the Borel parameters in the
$s$ and $s^\prime$ channel, respectively, and the function ${\cal
U}_0(\alpha,\beta)$ is defined as
\begin{eqnarray*}\label{eq25}
{\cal U}_0(a,b) = \int_0^\infty dy (y+M_1^2+M_2^2)^a y^b
\,exp\left[ -\frac{B_{-1}}{y} - B_0 - (B_{+1}) y \right]~, \nnb
\end{eqnarray*}

where

\begin{eqnarray*}\label{eq26}
B_{-1} \es \frac{1}{M_1^2M_2^2} \left[m_{u}^2M_1^4+m_b^2 M_2^4 +
M_2^2M_1^2 (m_b^2+m_{u}^2
-q^2) \right] ~, \nnb \\
B_0 \es \frac{1}{M_1^2 M_2^2} \left[ (m_{u}^2+m_c^2) M_1^2 + M_2^2
(m_b^2+m_c^2)
\right] ~, \nnb \\
B_{+1} \es \frac{m_c^2}{M_1^2 M_2^2}~.
\end{eqnarray*}

Performing the double Borel transformations over the variables
$p^2$ and $p^{'2}$ on the physical parts of the correlation
functions and bare-loop diagrams and also equating two
representations of the correlation functions, the sum rules for
the form factors $f_i$ are obtained:
\begin{eqnarray}\label{eq27}
f_{i}^{\prime} \es  \frac{(m_b+m_c)}{f_{B_c} m_{B_c}^2 f_{D^{*}} m_{D^{*}}}
e^\frac{m_{B_c}^2}{M_1^2}
e^\frac{m_{D^*}^2}{M_2^2} \nnb \\
\cp\Bigg\{-\frac{1}{4 \pi^2} \int_{m_b^2}^{s_0^\prime} ds^\prime
\int_{s_L}^{s_0} \rho_{i} (s,s^\prime,q^2) e^\frac{-s}{M_1^2}
e^\frac{-s^{'}}{M_2^2} - iM^{2}_{1}M^{2}_{2}  \lla
\frac{\alpha_s}{\pi} G^2 \rra \frac{C_4^i}{6} \Bigg\}~.
\end{eqnarray}

where $i=V,0,+1$ and $2$, $s_0$ and $s_0^{'}$ are the continuum
thresholds in the pseudoscalar $B_c$ and vector $D^{*}$ channels,
respectively, and the lower bound integration limit of $s_{L}$ is as
follow:
\begin{eqnarray*}\label{eq28}
s_L \es \frac{(m_c^2+q^2-m_b^2-s^\prime) (m_b^2 s^\prime -m_c^2
q^2)} {(m_b^2-q^2) (m_c^2-s^\prime)}~.
\end{eqnarray*}
The explicit expressions for the $C_4^{i}$ are presented in the
Appendix--A.

\section{Heavy quark effective theory}

In the present section, we analyze the infinite heavy quark mass limit of the form
factors of the $B_c\rightarrow D^{*}\ell\nu$ calculated by 3PSR. To
obtain the dependency of the form factors $f_V, f_0, f_1$ and
$f_2$ on $y$ the following parameterization is used: (see
also\cite{31})
\begin{equation}\label{eq29}
y=\nu\nu'=\frac{m_{B_{c}}^2+m_{D^{*}}^2-q^2}{2m_{B_{c}}m_{D^*}}
\end{equation}

We also apply these definitions:
\begin{eqnarray}\label{eq30}
m_{c} &=&\frac{m_{b}}{\sqrt{Z}}, \nonumber\\
\sqrt{Z} &=&y+\sqrt{y^{2}-1},\nonumber \\
T_{1} &=&\frac{M_{1}^{2}}{2m_{b}},\nonumber \\
T_{2} &=&\frac{M_{2}^{2}}{2m_{c}}.
\end{eqnarray}
Where $m_b\rightarrow \infty$. In the above expressions, $T_1$ and $T_2$
are the new Borel parameters. The mass of light quark $u$ is set to
zero.

The continuum thresholds $\nu_{0}$, $\nu_{0}^{'}$ and integration
variables $\nu$, $\nu^{'}$ are defined as:
\begin{equation}\label{eq31}
\nu_{0}=\frac{s_{0}-m_{b}^2}{m_{b}},~~~~~~
\nu'_{0}=\frac{s'_{0}-m_{c}^2}{m_{c}},
\end{equation}
\begin{equation}\label{eq32}
\nu=\frac{s-m_{b}^2}{m_{b}},~~~~~~ \nu'=\frac{s'-m_{c}^2}{m_{c}}.
\end{equation}

The leptonic decay constants are also rescaled  as:
\begin{equation}\label{eq33}
\hat{f}_{B_{c}}=\sqrt{m_{b}}
f_{B_{c}},~~~~~~~\hat{f}_{D^{*}}=\sqrt{m_{c}} f_{D^{*}}.
\end{equation}

The correspond expressions for $\overline{I}_0(a,b,c)$,
$\overline{I}_{1(2)}(a,b,c)$, $\overline{I}_j(a,b,c)$; $j=3,4,5$ and
$\overline{I}_6(a,b,c)$ in this limit are defined as:
\begin{eqnarray}  \label{eq34}
\overline{I}_0(a,b,c)\!\!\! &=& \!\!\! \frac{(-1)^{a+b+c}}{16
\pi^2\,\Gamma(a)
\Gamma(b) \Gamma(c)} (\frac{1}{\sqrt{Z}})^{2-a-c}(2m_b)^{4-2a-b-c}~T_1^{2-a-b} ~T_2^{2-a-c} \,\nonumber\\
&&\mathcal{U}%
_0^{HQET}(a+b+c-4,1-c-b)~,  \nonumber \\
\overline{I}_{1(2)}(a,b,c) \!\!\! &=& \!\!\! i
\frac{(-1)^{a+b+c+1}}{16 \pi^2\,\Gamma(a) \Gamma(b)
\Gamma(c)}(\frac{1}{\sqrt{Z}})^{4-a-c-1(2)}(2m_b)^{5-2a-b-c}
~T_1^{1-a-b+1(2)}~T_2^{4-a-c-1(2)} \,\nonumber\\
&&\mathcal{U}_0^{HQET}(a+b+c-5,1-c-b)~,  \nonumber \\
\nonumber \\
\overline{I}_j(a,b,c)\!\!\! &=& \!\!\! i \frac{(-1)^{a+b+c}}{16
\pi^2\,\Gamma(a) \Gamma(b) \Gamma(c)}
(\frac{1}{\sqrt{Z}})^{7-a-c-j}(2m_b)^{6-2a-b-c}~T_1^{-a-b-1+j}
~T_2^{7-a-c-j} \,\nonumber\\
&&\mathcal{U}_0^{HQET}(a+b+c-6,1-c-b)~,  \nonumber \\
\nonumber \\
\overline{I}_6(a,b,c)\!\!\! &=& \!\!\! i \frac{(-1)^{a+b+c+1}}{32
\pi^2\,\Gamma(a) \Gamma(b) \Gamma(c)}
(\frac{1}{\sqrt{Z}})^{3-a-c}(2m_b)^{6-2a-b-c}~T_1^{3-a-b}
~T_2^{3-a-c} \,\nonumber\\
&&\mathcal{U}_0^{HQET}(a+b+c-6,2-c-b)~.
\end{eqnarray}
The function $\mathcal{U}_0^{HQET}(m,n)$ takes the following form
\begin{eqnarray}\label{eq35}
\mathcal{U}_0^{HQET}(m,n)=\int _{0}^{\infty }\!
(2m_b)^{m}~(\frac{x}{2m_b}+T_1+\frac{T_2}{\sqrt{Z}})^{m}
~x^{n}[-\frac{\overline{B}_{-1}}{x}-\overline{B}_{0}-\overline{B}_{1}x]{dx},
\end{eqnarray}
with
\begin{eqnarray}\label{eq36}
\overline{B}_{-1}&=&\frac{\sqrt{Z}}{T_1T_2}~[\frac{mb^2}{Z}~T_2^2+\frac{1}{\sqrt{Z}}T_1T_2(m_b^2-q^2)],
\nonumber\\
\overline{B}_{0}&=&\frac{\sqrt{Z}}{2m_bT_1T_2}~[m_c^2T_1+\frac{T_2}{\sqrt{Z}}(m_b^2+m_c^2)],
\nonumber \\
\overline{B}_{1}&=&\frac{1}{4\sqrt{Z}T_1T_2}.
\end{eqnarray}

After some calculations, we obtain the y-dependent expressions of
the form factors as follows:
\begin{eqnarray}\label{eq37}
f_{V}^{HQET}(y) &=&\frac{1}{\hat{f}_{B_{c}}\hat{f}_{D^{*}}}e^{\frac{%
\Lambda }{T_{1}}}e^{\frac{\overline{{\Lambda
}}}{T_{2}}}\Bigg\{\frac{1}{{( 1+\sqrt {Z} )
{Z}^{\frac{7}{4}}\sqrt {{ \frac {-1+{y}^{2}}{Z}}}}}\nonumber\\
&&\Bigg[-3+ ( 3+9\,y ) \sqrt {Z}-6\,y ( 1+y ) Z  \Bigg] \nonumber\\
&&\frac{1}{(2\pi)^2}\int_{0}^{\nu _{0}}d\nu \int_{0}^{\nu
_{0}^{^{\prime }}}d{\nu }^{^{\prime }}e^{-\frac{\nu
}{2T_{1}}}e^{-\frac{{\nu }^{_{^{\prime }}}}{2T_{2}}}\theta
(2y\nu \nu ^{^{\prime }}-\nu ^{2}-{{\nu }^{_{^{\prime }}}}^{2})\nonumber \\
&&+\lim_{m_{b}\rightarrow \infty }~~\Bigg(i~\frac{Z^{\frac{3}{4}}}{3{m_{b}^{3}(1+\sqrt{Z})}}~%
\left\langle \frac{\alpha _{s}}{\pi }G^{2}\right\rangle
C_{V}^{HQET}\Bigg)\Bigg\}~,
\end{eqnarray}

\begin{eqnarray}\label{eq38}
f_{0}^{HQET}(y) &=&\frac{1}{\hat{f}_{B_{c}}\hat{f}_{D^{*}}}e^{\frac{%
\Lambda }{T_{1}}}e^{\frac{\overline{{\Lambda
}}}{T_{2}}}\Bigg\{\frac{1}{{2( 1+\sqrt {Z} )
^{3}{Z}^{\frac{7}{4}}\sqrt {{ \frac {-1+{y}^{2}}{Z}}}}}\nonumber\\
&&\Bigg[ 3+3\,\sqrt {Z} \{ -1+4\,y+2\, ( -1+y ) ( 1+3\,y ) \sqrt
{Z}-4\,y ( -1+y+{y}^{2} ) Z \}\Bigg] \nonumber\\
&&\frac{1}{(2\pi)^2}\int_{0}^{\nu _{0}}d\nu \int_{0}^{\nu
_{0}^{^{\prime }}}d{\nu }^{^{\prime }}e^{-\frac{\nu
}{2T_{1}}}e^{-\frac{{\nu }^{_{^{\prime }}}}{2T_{2}}}\theta
(2y\nu \nu ^{^{\prime }}-\nu ^{2}-{{\nu }^{_{^{\prime }}}}^{2})\nonumber \\
&&+\lim_{m_{b}\rightarrow \infty }~~\Bigg(i~\frac{Z^{\frac{7}{4}}}{6{m_{b}^{5}(1+\sqrt{Z})^{3}}}~%
\left\langle \frac{\alpha _{s}}{\pi }G^{2}\right\rangle
C_{0}^{HQET}\Bigg)\Bigg\}~,
\end{eqnarray}

\begin{eqnarray}\label{eq39}
f_{1}^{HQET}(y) &=&\frac{1}{\hat{f}_{B_{c}}\hat{f}_{D^{*}}}e^{\frac{%
\Lambda }{T_{1}}}e^{\frac{\overline{{\Lambda
}}}{T_{2}}}\Bigg\{\frac{1}{4(1+\sqrt{Z})({\frac{-1+{y}^{2}}{Z}})^{\frac{3}{2}}{Z}^{{\frac{15}{4}}}}\nonumber\\
&&\Bigg[9-9\,(1+5\,y)\sqrt{Z}+(-6+45\,y+78\,{y}^{2})Z \nonumber\\
&&-6\,(1+y[-3+y(11+9\,y)]){Z}^{\frac{3}{2}}+12\,y(1+y[-1+y(2+y)]){Z}^{2}\Bigg] \nonumber\\
&&\frac{1}{(2\pi)^2}\int_{0}^{\nu _{0}}d\nu \int_{0}^{\nu
_{0}^{^{\prime }}}d{\nu }^{^{\prime }}e^{-\frac{\nu
}{2T_{1}}}e^{-\frac{{\nu }^{_{^{\prime }}}}{2T_{2}}}\theta
(2y\nu \nu ^{^{\prime }}-\nu ^{2}-{{\nu }^{_{^{\prime }}}}^{2})\nonumber \\
&&+\lim_{m_{b}\rightarrow \infty }~~\Bigg(i~\frac{Z^{\frac{3}{4}}}{6{m_{b}^{3}(1+\sqrt{Z})}}~%
\left\langle \frac{\alpha _{s}}{\pi }G^{2}\right\rangle
C_{1}^{HQET}\Bigg)\Bigg\}~,
\end{eqnarray}

\begin{eqnarray}\label{eq40}
f_{2}^{HQET}(y) &=&\frac{1}{\hat{f}_{B_{c}}\hat{f}_{D^{*}}}e^{\frac{%
\Lambda }{T_{1}}}e^{\frac{\overline{{\Lambda
}}}{T_{2}}}\Bigg\{\frac{1}{4(1+\sqrt{Z})({\frac{-1+{y}^{2}}{Z}})^{\frac{3}{2}}{Z}^{{\frac{15}{4}}}}\nonumber\\
&&\Bigg[9-9\,(1+3\,y)\sqrt{Z}+9\,(2+y)(-1+2\,y)Z \nonumber\\
&&+6\,(1+y[5+(-1+y)y]){Z}^{\frac{3}{2}}-12\,y(1+y[-1+y(2+y)]){Z}^{2} \Bigg] \nonumber\\
&&\frac{1}{(2\pi)^2}\int_{0}^{\nu _{0}}d\nu \int_{0}^{\nu
_{0}^{^{\prime }}}d{\nu }^{^{\prime }}e^{-\frac{\nu
}{2T_{1}}}e^{-\frac{{\nu }^{_{^{\prime }}}}{2T_{2}}}\theta
(2y\nu \nu ^{^{\prime }}-\nu ^{2}-{{\nu }^{_{^{\prime }}}}^{2})\nonumber \\
&&+\lim_{m_{b}\rightarrow \infty }~~\Bigg(i~\frac{Z^{\frac{3}{4}}}{6{m_{b}^{3}(1+\sqrt{Z})}}~%
\left\langle \frac{\alpha _{s}}{\pi }G^{2}\right\rangle
C_{2}^{HQET}\Bigg)\Bigg\}~.
\end{eqnarray}
In the heavy quark limit expressions of the form factors, the
$\Lambda=m_{B_c}-m_{b}$ and $\bar{\Lambda}=m_{D_u^*}-m_c$ and the
explicit expressions of the coefficients $C_{i}^{HQET}$ are given in
the Appendix--B.

At the end of this section, we would like to present the 
differential decay width ${d\Gamma}/{dq^2}$ for the process
$B_c^-\rightarrow D^{*0}\ell \nu$ in terms of the form factors
as follow:
\begin{eqnarray}\label{eq41}
\frac{d\Gamma _{\pm }(B_{c}\rightarrow D^{*}\ell \nu)}{dq^2}
&=&\frac{G^{2}\left| V_{ub}\right| ^{2}}{192\pi
^{3}m^3_{B_c}}~q^{2}\lambda
^{1/2}(m^2_{B_c},m^2_{D^*},q^{2})\left|
H_{\pm }\right| ^{2}~,  \nonumber \\
&&  \nonumber \\
\frac{d\Gamma _{0}(B_{c}\rightarrow D^{*}\ell \nu)}{dq^2}
&=&\frac{G^{2}\left| V_{ub}\right| ^{2}}{192\pi ^{3}m^3_{B_c}}~
q^{2}\lambda ^{1/2}(m^2_{B_c},m^2_{D^*},q^{2})\left| H_{0}\right|
^{2}~,
\end{eqnarray}
\begin{eqnarray*}\label{eq42}
H_{\pm }(q^{2})
&=&(m_{B_c}+m_{D^*})f_{0}(q^{2})\mp \frac{%
\lambda ^{1/2}(m^2_{B_c},m^2_{D^*},q^{2})}{%
m_{B_c}+m_{D^*}}f_{V}(q^{2})~, \\
\\
H_{0}(q^{2}) &=&\frac{1}{2m_{D^*}\sqrt{q^2}}\Bigg[%
(m^2_{B_c}-m^2_{D^*}-q^{2})(m_{B_c}+m_{D^*}%
)f_{0}(q^{2}) -\frac{\lambda
(m^2_{B_c},m^2_{D^*},q^{2})}{%
m_{B_c}+m_{D^*}}f_{1}(q^{2})\Bigg]~.
\end{eqnarray*}

where $\pm, 0$ refers to the $D^*$ helicities.

\section{Numerical analysis}
The sum rules expressions of the form factors depict that the main
input parameters entering the expressions are gluon condensate,
elements of the CKM matrix $V_{ub}$, leptonic decay constants,
$f_{B_C}$ and $f_{D^*}$, Borel parameters $M_{1}^2$ and $M_{2}^2$,
as well as the continuum thresholds $s_{0}$ and $s'_{0}$. We choose
the values of the Gluon condensate, leptonic decay constants, CKM
matrix elements, quark and meson masses as:
$<\frac{\alpha_{s}}{\pi}G^{2}>=0.012~GeV^{4}$ \cite{33}, $\mid
V_{ub}\mid=0.0037$ \cite{32}, $f_{B_c}=0.35\pm0.025
~GeV$\cite{29,36}, $f_{D^*}=0.22\pm0.016~GeV$,
$m_{c}(\mu=m_c)=1.275\pm 0.015~GeV$, $m_{u}=(1.5-3)~MeV$,
$m_{b}=(4.7\pm0.01)~GeV$, $m_{D^*}=2.007~GeV$,
$m_{B_C}=6.258~GeV$\cite{37}, $\Lambda=0.62 ~GeV$\cite{34} and
$\overline{\Lambda}=0.86~GeV$\cite{35}.

The expressions for the form factors  contain also four auxiliary
parameters: Borel mass squares $M_{1}^2$ and $M_{2}^2$ and continuum
threshold $s_{0}$ and $s'_{0}$. These are mathematical objects, so
the physical quantities, form factors, should be independent of
them. The parameters $s_0$ and $s_0^\prime$, which are the continuum
thresholds of $B_c$ and $D^*$ mesons, respectively, are determined
from the conditions that guarantee the sum rules to have the best
stability in the allowed $M_1^2$ and $M_2^2$ region. The values of
the continuum thresholds calculated from the two--point QCD sum
rules are taken to be $s_0=(45-50)~GeV^2$ and
$s_0^\prime=(6-8)~GeV^2$ \cite{13,33,38}. The working regions for
$M_1^2$ and $M_2^2$ are determined by requiring that not only
contributions of the higher states and continuum are effectively
suppressed, but the gluon condensate contributions are small, which
guarantees that the contributions of higher dimensional operators
are small. Both conditions are satisfied in the  regions $10~GeV^2
\le M_1^2 \le 20~GeV^2$ and $4~GeV^2 \le M_2^2 \le 10~GeV^2$.

The dependence of the form factors $%
f_{V}, f_{0}, f_{1}$ and $f_{2}$ on $M_1^2$ and $M_2^2$ for $B_c
\rightarrow D^* \ell\nu$ are shown in Fig. 3. This
figure shows a good stability of the form factors with respect to
the Borel mass parameters in the working regions. Our numerical
analysis shows that   the contribution of the non-perturbative part
(the gluon condensate diagrams ) is about $5\%$ of the total and the
main contribution comes from the perturbative part of the form
factors. This means that the contribution of the higher dimension
operators is small and this guarantees the convergence of the sum
rules expression of the form factors and  those sum rules are
reliable.

\begin{figure}[th]

\vspace*{4.cm}
\begin{center}
\begin{picture}(160,35)
\put(20,-55){ \epsfxsize=7cm \epsfbox{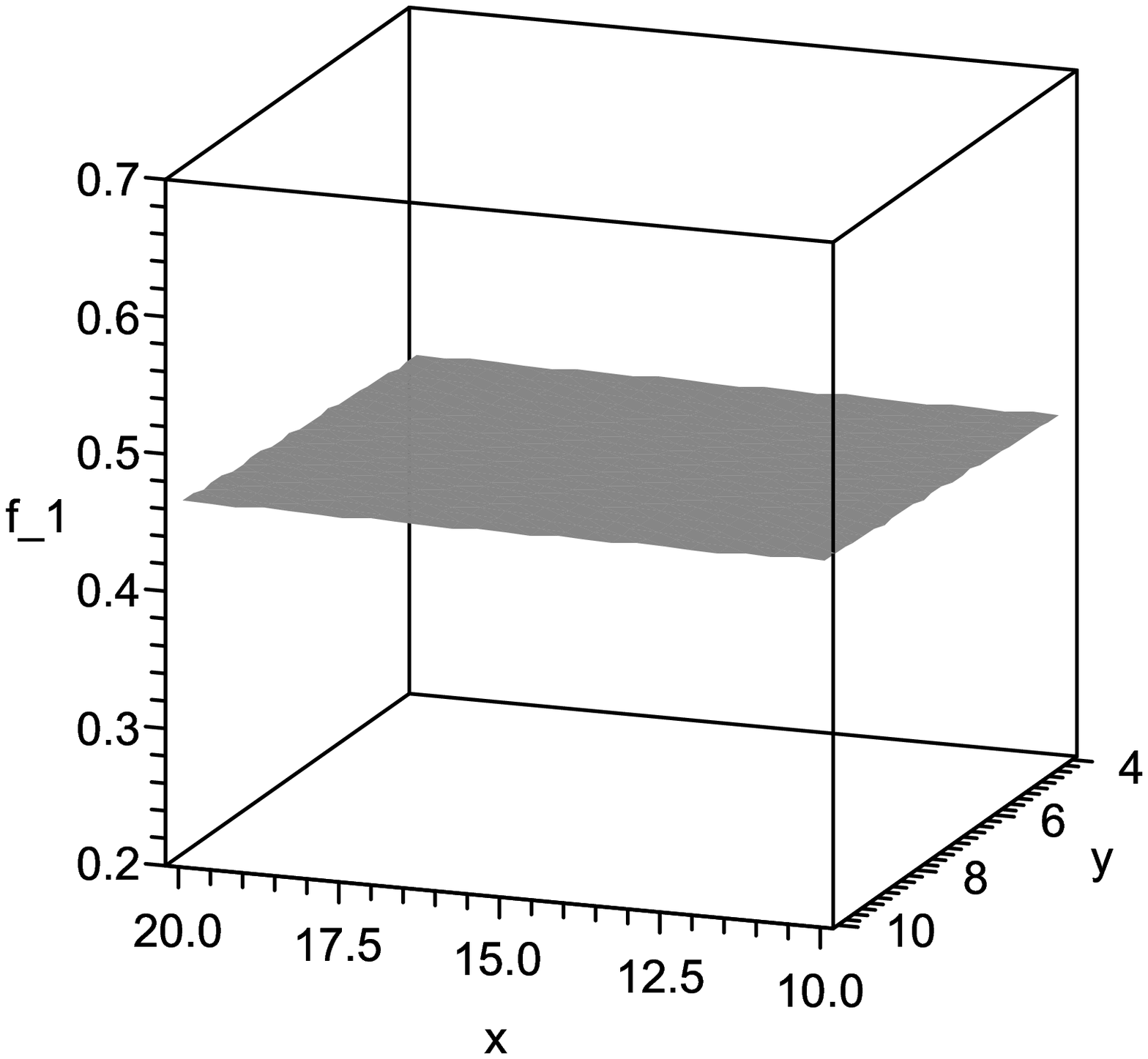}\epsfxsize=7cm
\epsfbox{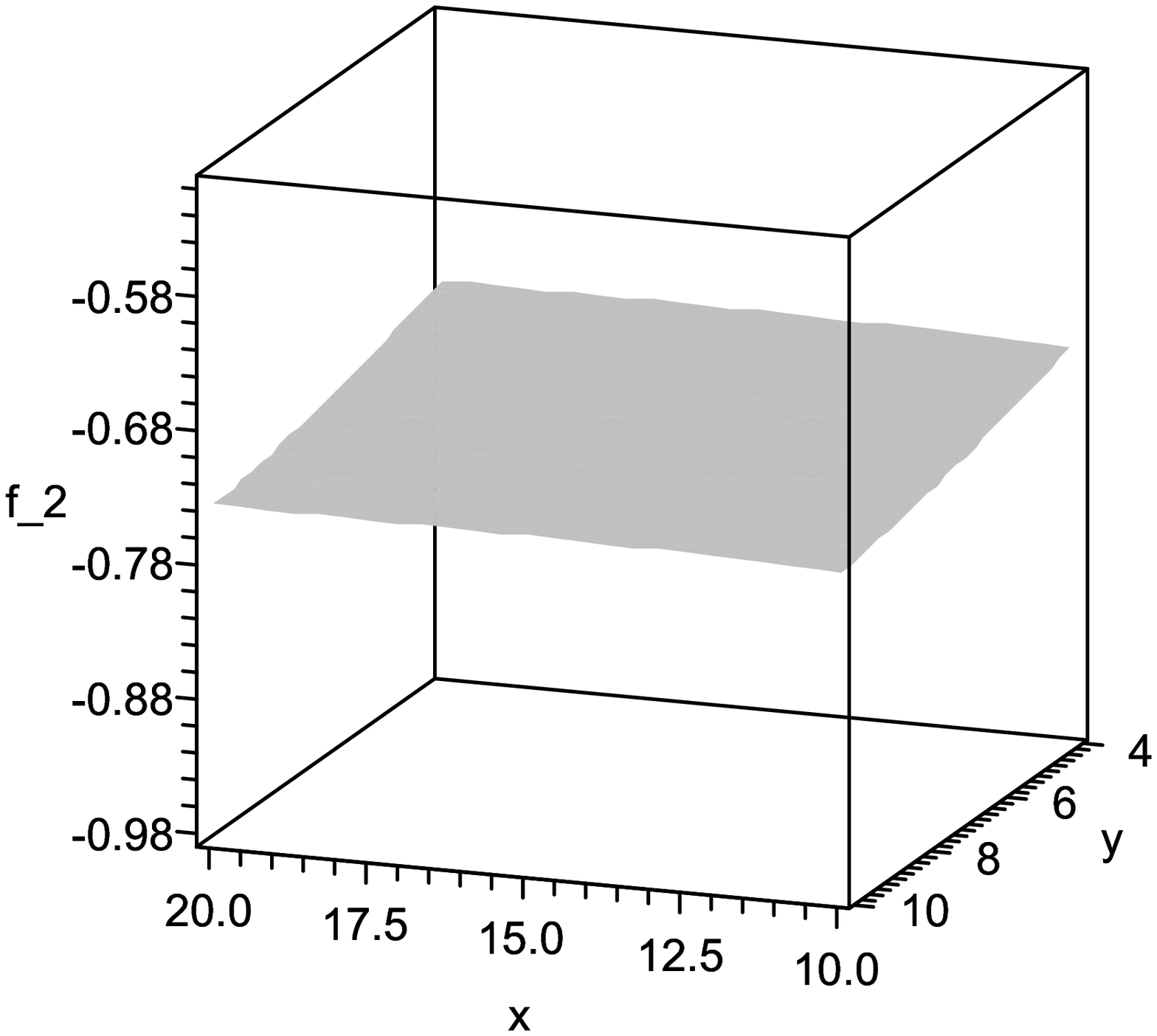}} \put(20,10){\epsfxsize=7cm
\epsfbox{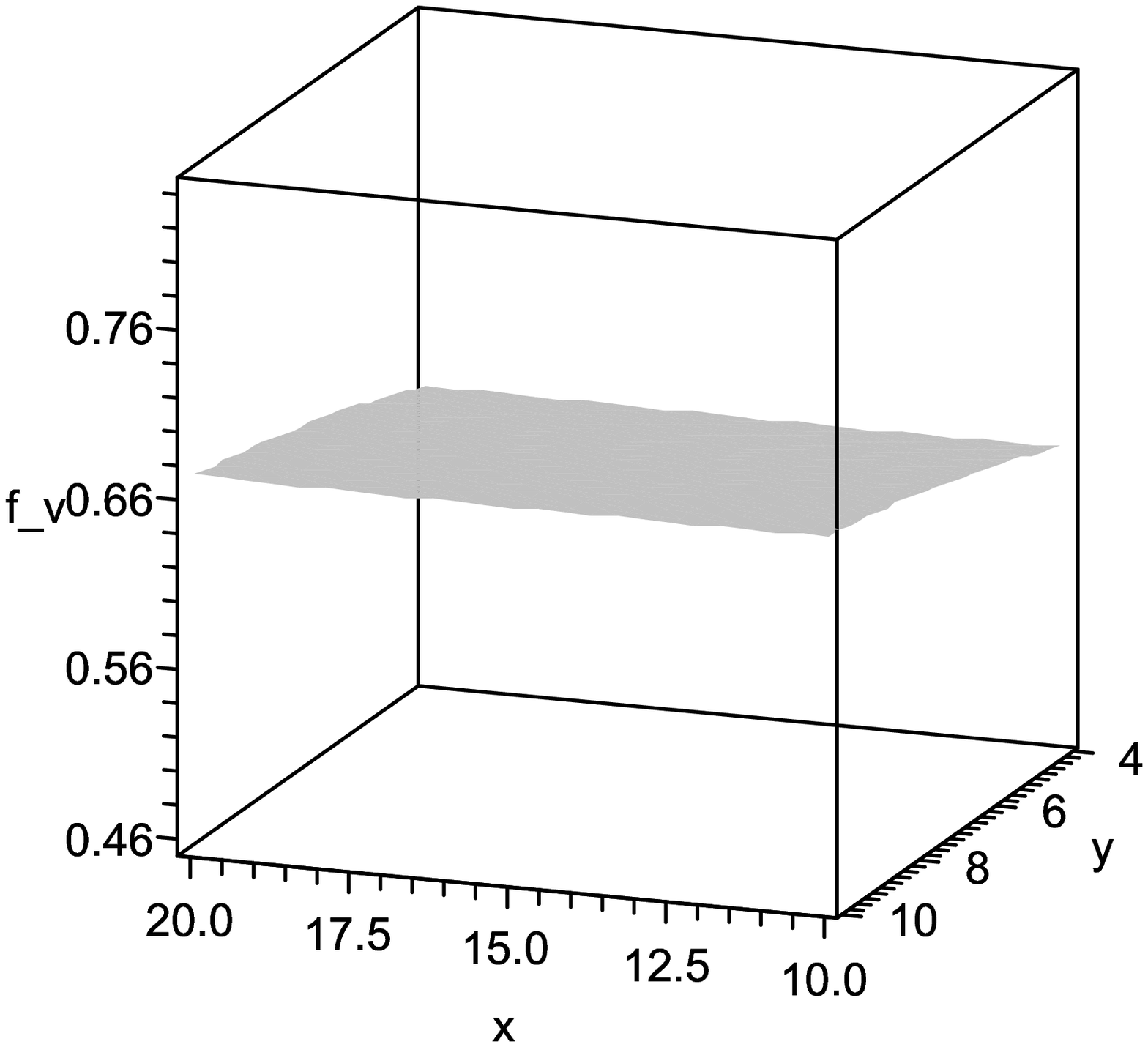}\epsfxsize=7cm \epsfbox{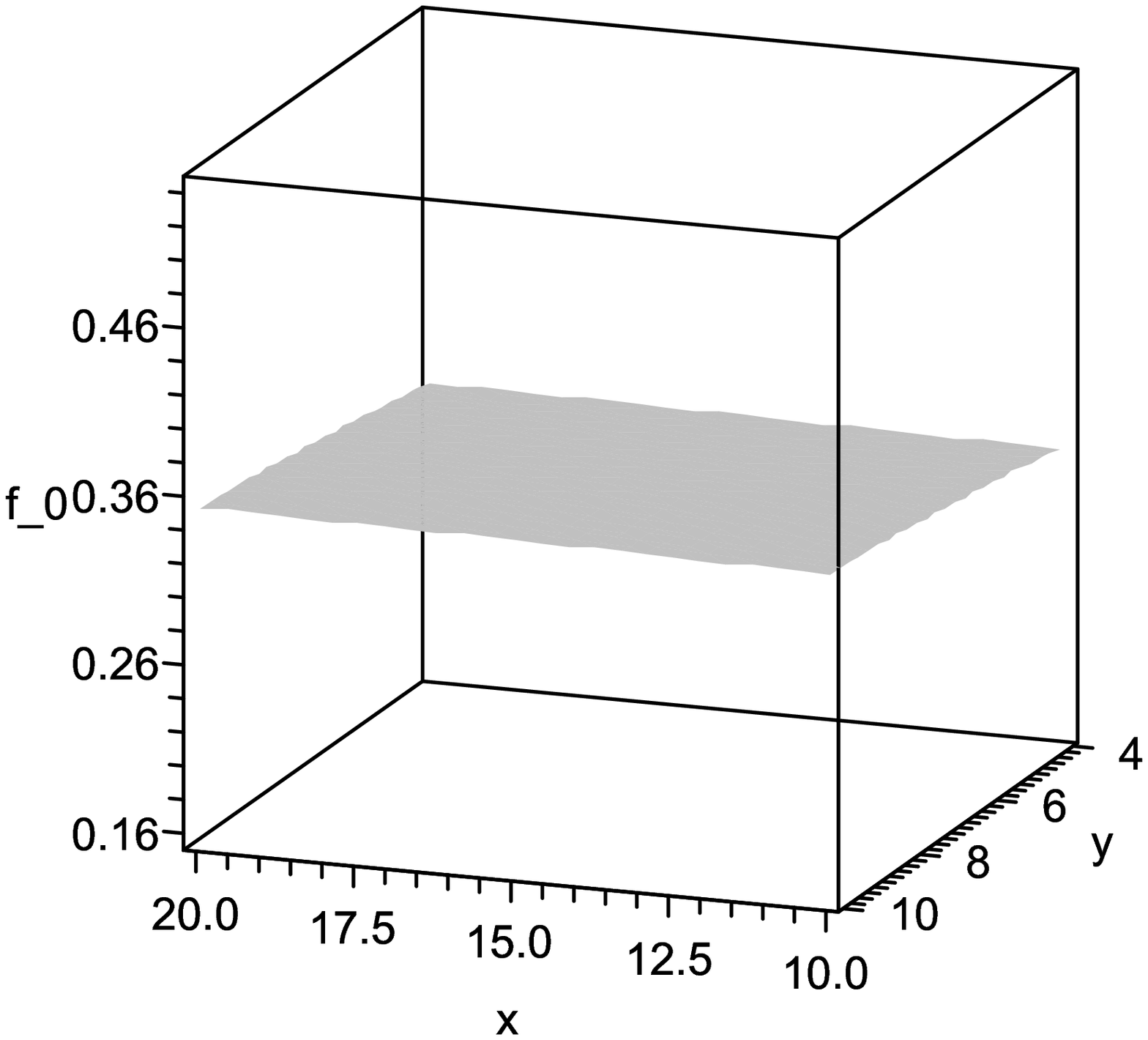}}
\end{picture}
\end{center}

\vspace*{5cm} \caption{The dependence of the form factors on
$x=M_1^2$ and $y=M_2^2$ for $B_c ^-\rightarrow D^{*0} \ell\nu$. }
\end{figure}
\normalsize

\setlength{\unitlength}{1mm}
\begin{table}[h]\label{tab:1}
\centering
\begin{tabular}{c|cc|c|c|c|c}

\cline{2-3}& Our&\\
\cline{1-7} Form factor & 3PSR & HQET&LCSR\cite{32}
&3PSR\cite{12}&PM\cite{18}& QM\cite{28}
\\ \cline{1-7}\lower0.35cm \hbox{{\vrule width 0pt height 1.0cm }}
$f^{B_c\rightarrow D^{*}}_{V}$ &$0.67\pm0.16$&$0.36\pm0.09$&0.57&0.83&0.80&0.98\\
\cline{1-7}\lower0.35cm \hbox{{\vrule width 0pt height 1.0cm }}
$f^{B_c\rightarrow D^{*}}_{0}$ &$0.35\pm0.09$&$0.25\pm0.06$&0.32&0.43&0.43&0.56 \\
\cline{1-7}\lower0.35cm \hbox{{\vrule width 0pt height 1.0cm }}
$f^{B_c\rightarrow D^{*}}_{1}$ &$0.46\pm0.11$&$0.20\pm0.05$&0.57&0.51&0.49&0.64  \\
\cline{1-7}\lower0.35cm \hbox{{\vrule width 0pt height 1.0cm }}
$f^{B_c\rightarrow D^{*}}_{2}$ &$-0.74\pm0.18$&$-0.35\pm0.09$&-0.57&-0.83&-0.89&-1.17  \\
\cline{1-7}
\end{tabular}
\vspace{0.8cm} \caption{The form factors of the $B_{c}^-\rightarrow
D^{*0}\ell\nu$ decay for $M_{1}^2=17~GeV^2$, $ M_{2}^2=8~GeV^2$ at
$q^2=0$ in different approaches: three--point sum rules (3PSR) with
gluon condensate corrections,  heavy quark effective theory (HQET),
 light cone sum
rules (LCSR), three--point sum rules without gluon condensate
corrections (3PSR), potential model(PM) and quark model(QM).}
\end{table}

The values of  the form factors at $q^2=0$ are  shown in Table.1. In
comparison, the predictions of the other approaches are also presented in
this Table.

The sum rules for the form factors are truncated at about $10
~GeV^2$, so to extend our results to the full physical region, we
look for a parameterization of the form factors in such a way that in
the region $0 \leq q^2 \leq 10~ GeV^2$, this parameterization
coincides with the sum rules predictions. Our numerical calculations
shows that the sufficient parameterization
of the form factors with respect to $q^2$ is as follows:%
\newline
\begin{equation}  \label{eq43}
f_{i}(q^2)=\frac{f_{i}(0)}{1+ \alpha\hat{q}+ \beta\hat{q}%
^2}
\end{equation}
where $\hat{q}=q^2/m_{B_{c}}^2$. The values of the parameters $%
f_{i}(0), ~\alpha$ and  $\beta$ are given in the Table 2.
\begin{table}[h]\label{tab:2}
\centering
\begin{tabular}{|c||c|c|c|}
\hline & f(0) & $ \alpha$ & $ \beta$
\\ \cline{1-4} \hline\hline$f_{V}$ &0.67&-0.53&-0.26\\
\cline{1-4} $f_{0}$ &0.35&0.38&-3.08\\
\cline{1-4} $f_{1}$ &0.46&1.92&-16.43\\
\cline{1-4}$f_{2}$ &-0.74&1.04&-15.77\\
\cline{1-4}
\end{tabular}
\vspace{0.8cm} \caption{Parameters appearing in the form factors
of the $B_{c}\rightarrow D^*\ell\nu$ decay  for $M_{1}^2=17
~GeV^2$, $M_{2}^2=8~GeV^2$.}
\end{table}

The errors are estimated by the variation of the Borel parameters
$M_1^2$
and $M_2^2$, the variation of the continuum thresholds $s_0$ and $s_0^\prime$%
, the leptonic decay constants $%
f_{B_c}$ and $f_{D^*}$ and uncertainties in the values of the other
input parameters. The main uncertainty comes from the continuum
thresholds and the decay constants, which is about $\sim 19\%$ of
the central value, while the other uncertainties are small,
constituting a few percent.

Now, we compare the values for the form factors and their HQET
values obtained from Eqs. (\ref{eq37}-\ref{eq40}) in Table 3 for
$B_{c}^-\rightarrow D^{*0} \ell \nu$.
\begin{table}[h]\label{tab:3}
\centering
\begin{tabular}{||c||ccccccccc||}
\hline $y$ &1&1.02& 1.1& 1.2& 1.3& 1.4& 1.5& 1.6& 1.7\\
\cline{1-10}$q^2$ &18.00&17.50 &15.49&12.99&10.48&7.97&5.46&2.95&0.44\\
\cline{1-10}\hline\hline$f_{V}(q^2)$ &5.81&5.11 &2.87 &1.86 &1.38&1.10 &0.91 &0.79 &0.68\\
\cline{1-10}\lower0.35cm \hbox{{\vrule width 0pt height 1.0cm }}
$f^{HQET}_{V}(y)$&-&4.98&0.84&0.59&0.50&0.48&0.43&0.39&0.37\\
\cline{1-10}\hline\hline$f_{0}(q^2)$ &0.83&0.72 &0.57&0.48&0.42&0.39&0.37&0.36&0.35\\
\cline{1-10}\lower0.35cm \hbox{{\vrule width 0pt height 1.0cm }}
$f^{HQET}_{0}(y)$ &-&0.69&0.36&0.32&0.30&0.29&0.28&0.27&0.26\\
\cline{1-10}\hline\hline$f_{1}(q^2)$  &2.18&1.89 &1.19&0.84&0.68&0.59&0.52&0.49&0.47\\
\cline{1-10}\lower0.35cm \hbox{{\vrule width 0pt height 1.0cm }}
$f^{HQET}_{1}(y)$ &-&1.74 &0.35&0.29&0.26&0.24&0.23&0.22&0.21\\
\cline{1-10}\hline\hline$f_{2}(q^2)$ &-5.87&-5.01&-2.61 &-1.69 &-1.29 &-1.07&-0.96 & -0.85&-0.76\\
\cline{1-10}\lower0.35cm \hbox{{\vrule width 0pt height 1.0cm }}
$f^{HQET}_{2}(y)$ &-&-4.79 &-0.62 &-0.49 &-0.44 &-0.42 &-0.39 & -0.37&-0.36\\
\cline{1-10}
\end{tabular}
\vspace{0.10cm} \caption{The comparison of the values for the form
factors and their HQET limit for $B_{c}^-\rightarrow D^{*0} \ell
\nu$ at $M_1^{2}=17~GeV^2, M_2^{2}=8~GeV^2$ and corresponding
$T_1=1.80 ~GeV, T_2=3.14~GeV$.}
\end{table}

At  $y=1$ called the zero recoil limit, the HQET limit of the form
factors are not finite and at this value, we can determine only the
ratio of the form factors. For other values of y and corresponding
$q^2$, the behavior of the form factors and their HQET values are
the same, i.e., when y increases ($q^2$ decreases) both the form
factors and their HQET values decrease. Moreover, at high $q^2$
values, the form factors and their HQET values are close to each
other. For better comparison we prefer to plot the dependence of the
relevant form factors and HQET limit of them on the momentum
transfer square $q^2$ ( Fig. 4). This figure shows a good agreement
 between both  form factors and their HQET at high $q^2$ values.
This figure also contains  the fit function of the form factors (see
Eq. (\ref{eq27})). The form factors and their fit functions coincide
well in the interval $0 \leq q^2 \leq 10~ GeV^2$.

At the end of this section, we would like to present the values of
the branching ratio for $B_{c}^-\rightarrow D^{*0} \ell
\nu$. Integrating Eq. (\ref{eq41}) over $q^2$ in
the whole physical region and using the total mean life time $\tau
\simeq 0.48\pm0.05~ps$ of $B_c$ meson \cite {39}, the branching
ratio of the $B_c^- \rightarrow D^{*0}\ell\nu$ decay is obtained as
presented in Table 4. The branching ratio  of this decay obtained
using the HQET limit of form factors Eqs. (\ref{eq37}-\ref{eq40}) is
also shown in this Table. This Table also includes a comparison
between our results via both SR and HQET and the predictions of the
other approaches including the LCSR, 3PSR (without gluon condensate
corrections), QM, BSE, PM and RM estimates.

\begin{table}[h]\label{tab:4}
\centering
\begin{tabular}{c|cc|c|c|c|c|c|c}
\cline{2-3} &Our&\\
\hline Mode & 3PSR&HQET&  LCSR\cite{32} & 3PSR\cite{12} &
QM\cite{28} &PM\cite{24}&BSE\cite{30}&RM\cite{21}
\\ \cline{1-9} \hline\hline$D^* e \nu$ &$(2.2\pm0.5)10^{-2}$&$(3\pm0.7)10^{-3}$&0.035&0.018&0.034&0.004&0.018&0.013\\
\cline{1-9} $D^* \tau \nu$ &$(1.2\pm0.3)10^{-2}$&$(1\pm0.2)10^{-3}$&0.020&0.008&0.019&-&-&-\\
\cline{1-9}
\end{tabular}
\vspace{0.8cm} \caption{The branching ratio of the
$B_{c}^-\rightarrow D^{*0}\ell\nu$ decay    in different approaches: 3PSR with gluon
condensate corrections, HQET,  LCSR \cite{32}, 3PSR without gluon
corrections \cite{12} , QM \cite{28}, PM \cite{24}, the
Bethe-Salpeter equation(BSE) \cite{30} and a relativistic
model with factorization to obtain the nonleptonic decay widths (RM) \cite{21}.}
\end{table}

\section*{Conclusion }
Considering the gluon corrections, we investigated the $B_c^-
\rightarrow D^{*0} \ell\nu$ channel in the frame work of
three--point QCD sum rules. We found that the gluon correction
contributions to the sum rules expression of the form factors are
small. This implies the  small contribution of the higher dimension
operators and also it guarantees that the sum rules for the form
factors are convergent and reliable. The HQET limit of the form
factors with their corresponding gluon condensate corrections are
also computed. A Comparison between the form factors and their HQET
 was made. Finally, we evaluated the total decay width and the
branching fraction of this decay and compared  with the predictions
of the other approaches  such as LCSR, 3PSR, QM, PM, BSE and RM.

\begin{figure}[th]

\vspace*{4.cm}
\begin{center}
\begin{picture}(160,100)
\put(7,0){ \epsfxsize=7cm \epsfbox{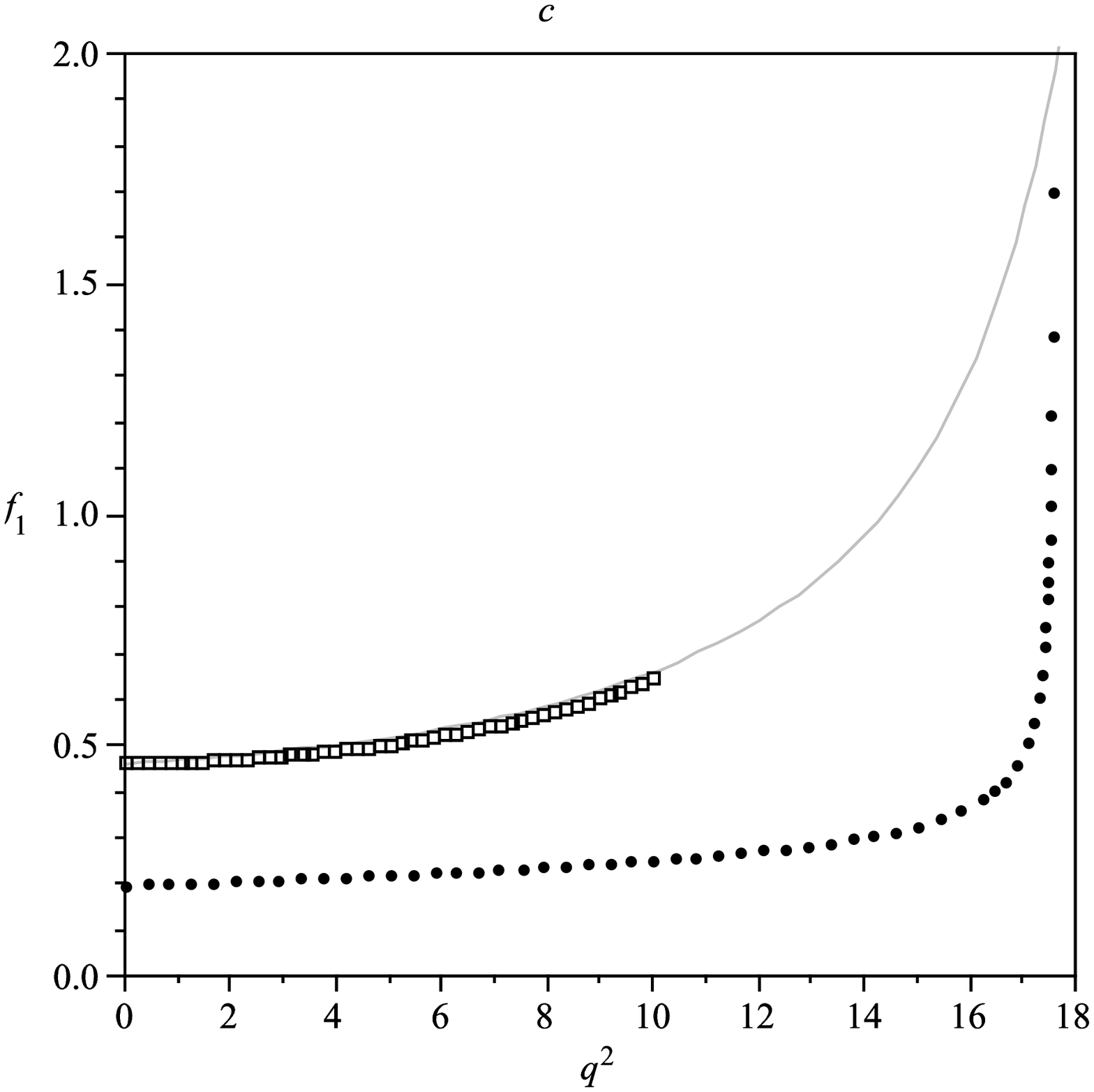}\epsfxsize=7cm
\epsfbox{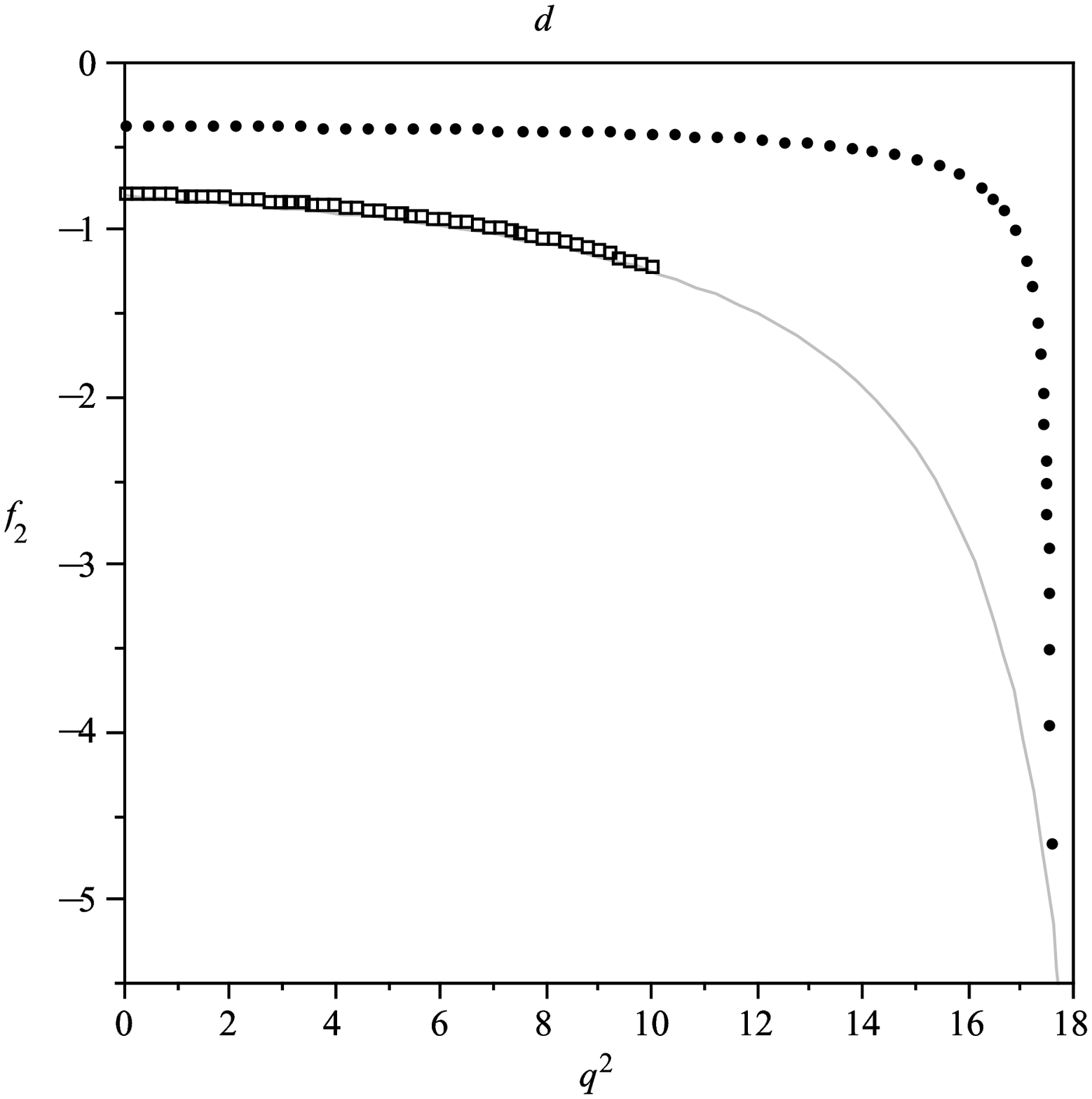}} \put(9,70){\epsfxsize=7cm
\epsfbox{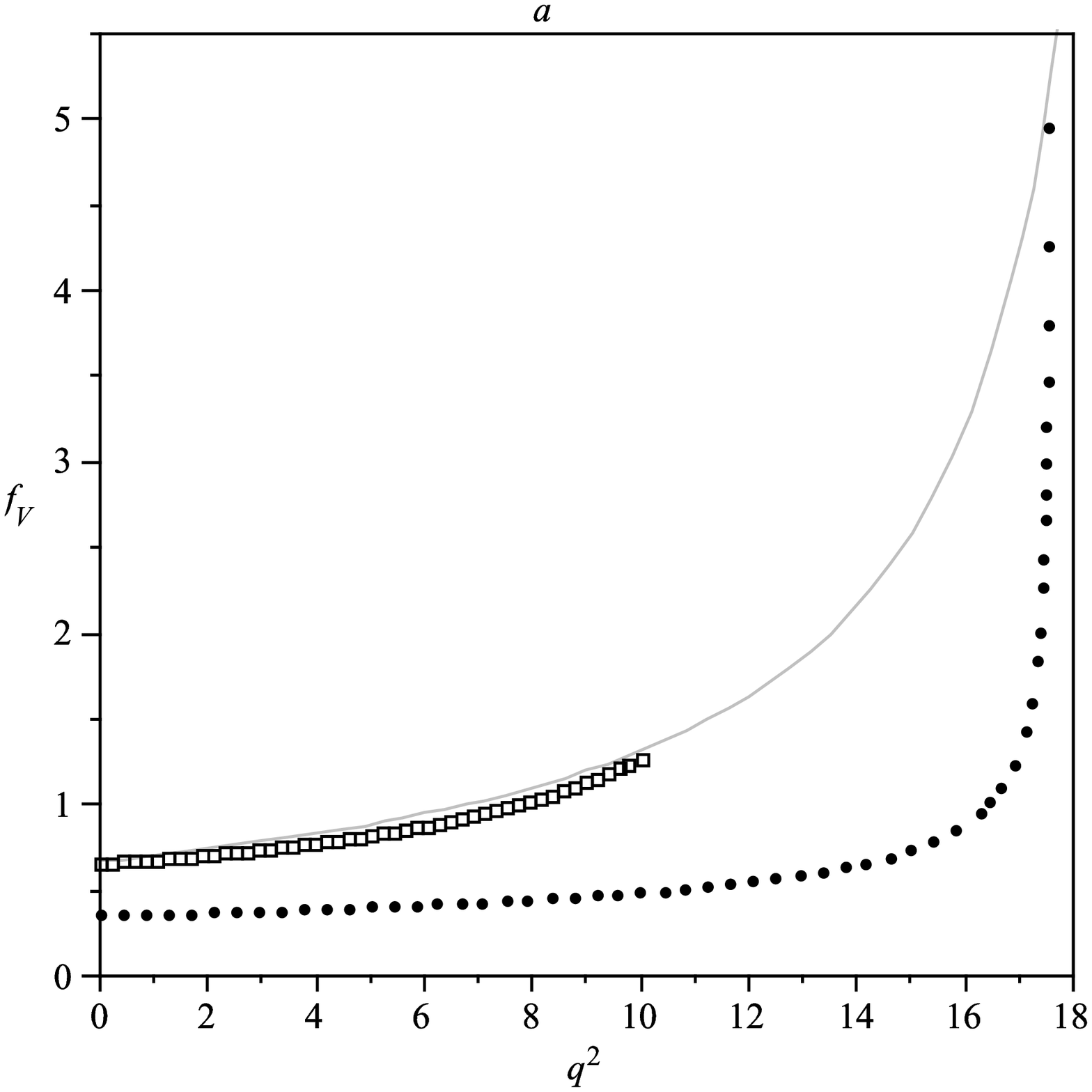}\epsfxsize=7cm \epsfbox{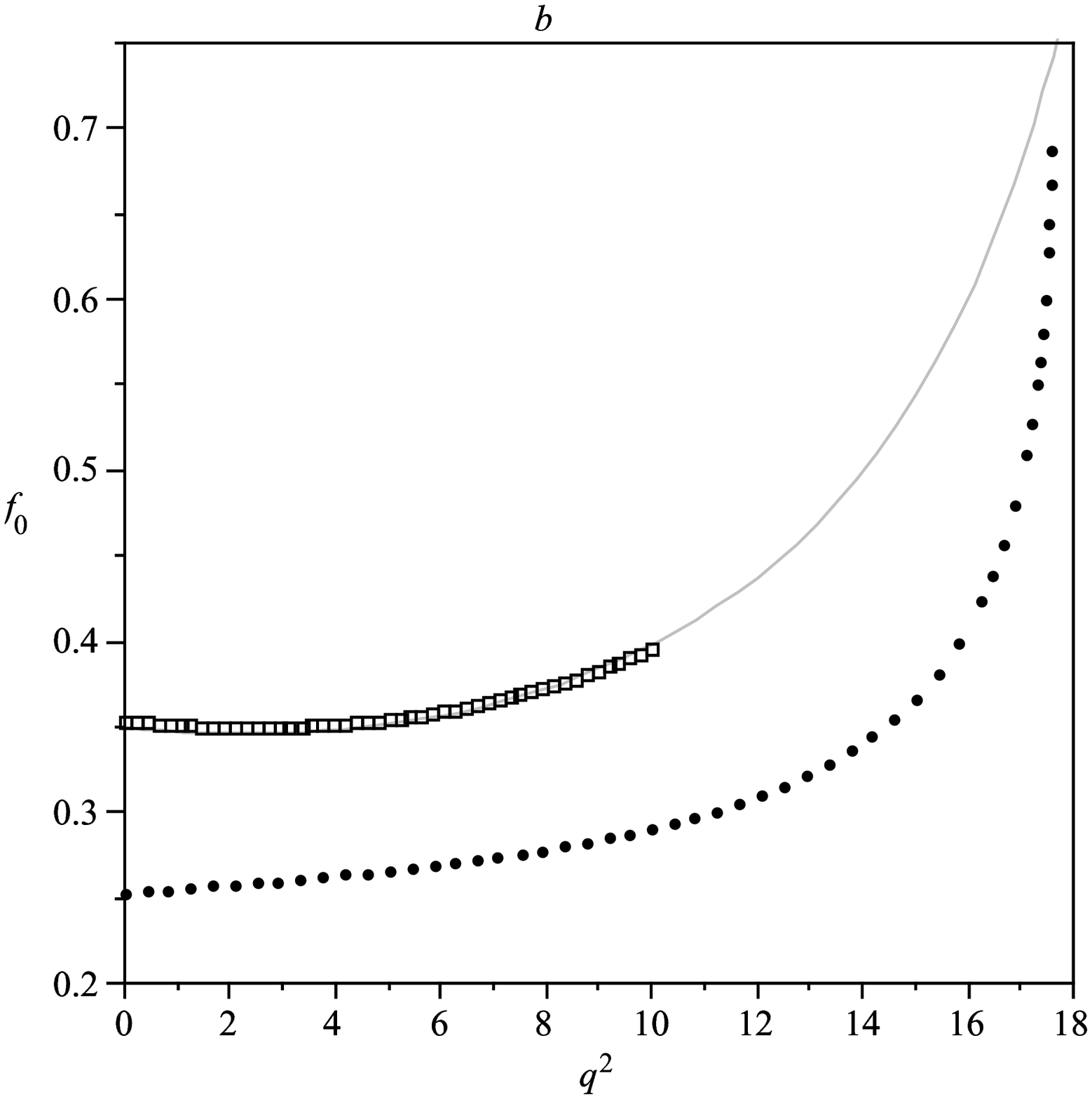}}
\end{picture}
\end{center}

\vspace*{-1cm} \caption{The dependence of the form factors and their
HQET limit  as well as the fit parameterization of the form factors
 on $q^2$. The small boxes correspond to the form factors, the solid lines are belong to the fit parameterization of  form factors and dotted lines show
  the HQET limit of the form factors.}
\end{figure}
\normalsize

\setlength{\unitlength}{1mm}

\newpage
\section*{}
\newpage

\section*{Acknowledgments}
Partial support of Shiraz university research council is
appreciated. K. Azizi would like to thank T. M. Aliev and A.
Ozpineci for their usefull discussions and also TUBITAK, turkish
scientific and research  council, for their partial financial
support.
\newpage

\appendix

\begin{center} {\Large \textbf{Appendix--A}}
\end{center}

\setcounter{equation}{0} \renewcommand{\theequation}

In this appendix, the explicit expressions of the coefficients of
the gluon condensate  entering  the sum rules of the form factors
$f_{V}, f_{0}, f_{1}$ and $f_{2}$ are given.

\begin{eqnarray*}
C_4^V &=&-10\,\hat{I}_{{1}}(3,2,2){{m_{c}}}^{5}-10\,\hat{I}_{{2}}(3,2,2){{m_{c}}}%
^{5}-10\,\hat{I}_{{0}}(3,2,2){{m_{c}}}^{5}+10\,\hat{I}_{{1}}(3,2,2){{m_{c}}}^{4}{m_{b}}
\\
&&+10\,\hat{I}_{{0}}(3,2,2){{m_{c}}}^{3}{{m_{b}}}^{2}+10\,\hat{I}_{{2}}(3,2,2){{m_{c}}}%
^{3}{{m_{b}}}^{2}+10\,\hat{I}_{{1}}(3,2,2){{m_{c}}}^{3}{{m_{b}}}^{2}-10\,\hat{I}_{{1}%
}(3,2,2){{m_{c}}}^{2}{{m_{b}}}^{3} \\
&&-10\,\hat{I}_{{0}}(3,1,2){{m_{c}}}^{3}+20\,\hat{I}_{0}^{[0,1]}(3,2,2){{m_{c}}}%
^{3}+20\,\hat{I}_{2}^{[0,1]}(3,2,2){{m_{c}}}^{3}-30\,\hat{I}_{{0}}(4,1,1){{m_{c}}}^{3} \\
&&-20\,\hat{I}_{{2}}(2,2,2){{m_{c}}}^{3}-10\,\hat{I}_{{2}}(3,1,2){{m_{c}}}^{3}+10\,\hat{I}_{{2}%
}(3,2,1){{m_{c}}}^{3}-20\,\hat{I}_{{0}}(2,2,2){{m_{c}}}^{3} \\
&&-20\,\hat{I}_{{1}}(2,2,2){{m_{c}}}^{3}-30\,\hat{I}_{{2}}(4,1,1){{m_{c}}}^{3}-30\,\hat{I}_{{1}%
}(4,1,1){{m_{c}}}^{3}+20\,\hat{I}_{1}^{[0,1]}(3,2,2){{m_{c}}}^{3} \\
&&+30\,\hat{I}_{{1}}(4,1,1){{m_{c}}}^{2}{m_{b}}+20\,\hat{I}_{{1}}(2,2,2){{m_{c}}}^{2}{%
m_{b}}-10\,\hat{I}_{{2}}(3,2,1){{m_{c}}}^{2}{m_{b}}-20\,\hat{I}_{1}^{[0,1]}(3,2,2){{m_{c}%
}}^{2}{m_{b}} \\
&&+10\,\hat{I}_{{1}}(3,2,1){{m_{c}}}^{2}{m_{b}}+20\,\hat{I}_{{1}}(2,3,1){{m_{c}}}^{2}{%
m_{b}}+40\,\hat{I}_{{0}}(2,3,1){{m_{c}}}^{2}{m_{b}}+10\,\hat{I}_{{0}}(3,2,1){{m_{c}}}^{2}%
{m_{b}} \\
&&+10\,\hat{I}_{2}^{[0,1]}(3,2,2){m_{c}}\,{{m_{b}}}^{2}+10\,\hat{I}_{1}^{[0,1]}(3,2,2){%
m_{c}}\,{{m_{b}}}^{2}-30\,\hat{I}_{{1}}(3,2,1){m_{c}}\,{{m_{b}}}^{2}-20\,\hat{I}_{{2}%
}(3,2,1){m_{c}}\,{{m_{b}}}^{2} \\
&&-20\,\hat{I}_{{0}}(3,2,1){m_{c}}\,{{m_{b}}}^{2}+10\,\hat{I}_{0}^{[0,1]}(3,2,2){m_{c}}\,%
{{m_{b}}}^{2}+60\,\hat{I}_{{1}}(1,4,1){m_{c}}\,{{m_{b}}}^{2}+60\,\hat{I}_{{0}}(1,4,1){%
m_{c}}\,{{m_{b}}}^{2} \\
&&-20\,\hat{I}_{{1}}(2,3,1){{m_{b}}}^{3}-60\,\hat{I}_{{1}}(1,4,1){{m_{b}}}%
^{3}-10\,\hat{I}_{1}^{[0,1]}(3,2,2){{m_{b}}}^{3}+10\,\hat{I}_{{1}}(2,2,2){{m_{b}}}^{3} \\
&&+20\,\hat{I}_{{1}}(3,2,1){{m_{b}}}^{3}+20\,\hat{I}_{{1}}(2,2,1){m_{c}}%
+30\,\hat{I}_{2}^{[0,1]}(3,1,2){m_{c}}-30\,\hat{I}_{{0}}(2,1,2){m_{c}} \\
&&-20\,\hat{I}_{{1}}(1,2,2){m_{c}}+20\,\hat{I}_{2}^{[0,1]}(2,2,2){m_{c}}%
-10\,\hat{I}_{1}^{[0,2]}(3,2,2){m_{c}}-20\,\hat{I}_{{2}}(1,2,2){m_{c}} \\
&&+20\,\hat{I}_{1}^{[0,1]}(2,2,2){m_{c}}+10\,\hat{I}_{{2}}(3,1,1){m_{c}}-20\,\hat{I}_{{0}%
}(1,2,2){m_{c}}-10\,\hat{I}_{{1}}(3,1,1){m_{c}} \\
&&+20\,\hat{I}_{1}^{[0,1]}(3,2,1){m_{c}}+20\,\hat{I}_{0}^{[0,1]}(3,2,1){m_{c}}+30\,\hat{I}_{{2}%
}(2,2,1){m_{c}}-10\,\hat{I}_{0}^{[0,2]}(3,2,2){m_{c}} \\
&&+20\,\hat{I}_{{0}}(2,2,1){m_{c}}+30\,\hat{I}_{0}^{[0,1]}(3,1,2){m_{c}}%
-10\,\hat{I}_{2}^{[0,2]}(3,2,2){m_{c}}+20\,\hat{I}_{1}^{[0,1]}(3,1,2){m_{c}} \\
&&+20\,\hat{I}_{0}^{[0,1]}(2,2,2){m_{c}}-30\,\hat{I}_{{2}}(2,1,2){m_{c}}-10\,\hat{I}_{{0}%
}(3,1,1){m_{c}}+10\,\hat{I}_{2}^{[0,1]}(3,2,1){m_{c}} \\
&&-20\,\hat{I}_{{1}}(2,1,2){m_{c}}+20\,\hat{I}_{1}^{[0,1]}(2,3,1){m_{b}}+30\,\hat{I}_{{1}%
}(2,1,2){m_{b}}+10\,\hat{I}_{1}^{[0,2]}(3,2,2){m_{b}} \\
&&+40\,\hat{I}_{2}^{[0,1]}(2,3,1){m_{b}}+20\,\hat{I}_{{1}}(1,2,2){m_{b}}-20\,\hat{I}_{{0}%
}(2,2,1){m_{b}}+60\,\hat{I}_{{2}}(1,3,1){m_{b}} \\
&&-20\,\hat{I}_{1}^{[0,1]}(3,2,1){m_{b}}-20\,\hat{I}_{1}^{[0,1]}(2,2,2){m_{b}}+100\,\hat{I}_{{0%
}}(1,3,1){m_{b}}-20\,\hat{I}_{1}^{[0,1]}(3,1,2){m_{b}} \\
&&-50\,\hat{I}_{{1}}(2,2,1){m_{b}}-20\,\hat{I}_{{2}}(2,2,1){m_{b}}+20\,\hat{I}_{{1}}(1,3,1){%
m_{b}}
\end{eqnarray*}
\begin{eqnarray*}
C_4^0 &=&5\,\hat{I}_{{0}}(3,2,2){{m_{c}}}^{6}{m_{b}}-5\,\hat{I}_{{0}}(3,2,2){{m_{c}}}^{5}{{%
m_{b}}}^{2}-5\,\hat{I}_{{0}}(3,2,2){{m_{c}}}^{4}{{m_{b}}}^{3}+5\,\hat{I}_{{0}}(3,2,2){{%
m_{c}}}^{3}{{m_{b}}}^{4} \\
&&+20\,\hat{I}_{{6}}(3,2,2){{m_{c}}}^{5}-20\,\hat{I}_{{6}}(3,2,2){{m_{c}}}^{4}{m_{b}}%
+5\,\hat{I}_{{0}}(3,2,1){{m_{c}}}^{4}{m_{b}}+15\,\hat{I}_{{0}}(2,2,2){{m_{c}}}^{4}{m_{b}}
\\
&&-15\,\hat{I}_{0}^{[0,1]}(3,2,2){{m_{c}}}^{4}{m_{b}}+15\,\hat{I}_{{0}}(4,1,1){{m_{c}}}%
^{4}{m_{b}}+10\,\hat{I}_{{0}}(2,3,1){{m_{c}}}^{4}{m_{b}}-10\,\hat{I}_{{0}}(2,2,2){{m_{c}}%
}^{3}{{m_{b}}}^{2} \\
&&-5\,\hat{I}_{{0}}(3,2,1){{m_{c}}}^{3}{{m_{b}}}^{2}-20\,\hat{I}_{{6}}(3,2,2){{m_{c}}}%
^{3}{{m_{b}}}^{2}-15\,\hat{I}_{{0}}(4,1,1){{m_{c}}}^{3}{{m_{b}}}%
^{2}+10\,\hat{I}_{0}^{[0,1]}(3,2,2){{m_{c}}}^{3}{{m_{b}}}^{2} \\
&&+10\,\hat{I}_{{0}}(3,2,1){{m_{c}}}^{2}{{m_{b}}}^{3}-30\,\hat{I}_{{0}}(1,4,1){{m_{c}}}%
^{2}{{m_{b}}}^{3}-10\,\hat{I}_{{0}}(2,3,1){{m_{c}}}^{2}{{m_{b}}}^{3}+20\,\hat{I}_{{6}%
}(3,2,2){{m_{c}}}^{2}{{m_{b}}}^{3} \\
&&-10\,\hat{I}_{{0}}(3,2,1){m_{c}}\,{{m_{b}}}^{4}+5\,\hat{I}_{0}^{[0,1]}(3,2,2){m_{c}}\,{%
{m_{b}}}^{4}+30\,\hat{I}_{{0}}(1,4,1){m_{c}}\,{{m_{b}}}^{4}+40\,\hat{I}_{{6}}(3,2,1){{%
m_{c}}}^{3} \\
&&-5\,\hat{I}_{{0}}(3,1,1){{m_{c}}}^{3}+20\,\hat{I}_{{6}}(3,1,2){{m_{c}}}%
^{3}-40\,\hat{I}_{6}^{[0,1]}(3,2,2){{m_{c}}}^{3}+40\,\hat{I}_{{6}}(2,2,2){{m_{c}}}^{3} \\
&&+60\,\hat{I}_{{6}}(4,1,1){{m_{c}}}^{3}-40\,\hat{I}_{{6}}(3,2,1){{m_{c}}}^{2}{m_{b}}%
-20\,\hat{I}_{{6}}(3,1,2){{m_{c}}}^{2}{m_{b}}-20\,\hat{I}_{{6}}(2,2,2){{m_{c}}}^{2}{m_{b}%
} \\
&&-15\,\hat{I}_{0}^{[0,1]}(4,1,1){{m_{c}}}^{2}{m_{b}}-30\,\hat{I}_{0}^{[0,1]}(2,2,2){{%
m_{c}}}^{2}{m_{b}}-20\,\hat{I}_{0}^{[0,1]}(2,3,1){{m_{c}}}^{2}{m_{b}}%
-20\,\hat{I}_{0}^{[0,1]}(3,2,1){{m_{c}}}^{2}{m_{b}} \\
&&+10\,\hat{I}_{{0}}(3,1,1){{m_{c}}}^{2}{m_{b}}-60\,\hat{I}_{{6}}(4,1,1){{m_{c}}}^{2}{%
m_{b}}+15\,\hat{I}_{{0}}(2,1,2){{m_{c}}}^{2}{m_{b}}+40\,\hat{I}_{6}^{[0,1]}(3,2,2){{m_{c}%
}}^{2}{m_{b}} \\
&&+40\,\hat{I}_{{6}}(2,3,1){{m_{c}}}^{2}{m_{b}}+15\,\hat{I}_{0}^{[0,2]}(3,2,2){{m_{c}}}%
^{2}{m_{b}}-10\,\hat{I}_{0}^{[0,1]}(3,1,2){{m_{c}}}^{2}{m_{b}}-10\,\hat{I}_{{0}}(1,3,1){{%
m_{c}}}^{2}{m_{b}} \\
&&-15\,\hat{I}_{{0}}(2,2,1){{m_{c}}}^{2}{m_{b}}+20\,\hat{I}_{{0}}(1,2,2){{m_{c}}}^{2}{%
m_{b}}-120\,\hat{I}_{{6}}(1,4,1){m_{c}}\,{{m_{b}}}^{2}+30\,\hat{I}_{{0}}(1,3,1){m_{c}}\,{%
{m_{b}}}^{2} \\
&&+20\,\hat{I}_{{6}}(2,2,2){m_{c}}\,{{m_{b}}}^{2}+20\,\hat{I}_{0}^{[0,1]}(3,1,2){m_{c}}\,%
{{m_{b}}}^{2}+5\,\hat{I}_{0}^{[0,1]}(3,2,1){m_{c}}\,{{m_{b}}}^{2}+10%
\,\hat{I}_{0}^{[0,1]}(2,2,2){m_{c}}\,{{m_{b}}}^{2} \\
&&-5\,\hat{I}_{0}^{[0,2]}(3,2,2){m_{c}}\,{{m_{b}}}^{2}-20\,\hat{I}_{6}^{[0,1]}(3,2,2){%
m_{c}}\,{{m_{b}}}^{2}+20\,\hat{I}_{{6}}(3,2,1){m_{c}}\,{{m_{b}}}^{2}-15\,\hat{I}_{{0}%
}(2,1,2){m_{c}}\,{{m_{b}}}^{2} \\
&&+10\,\hat{I}_{{0}}(2,2,1){m_{c}}\,{{m_{b}}}^{2}-10\,\hat{I}_{{0}}(1,2,2){m_{c}}\,{{%
m_{b}}}^{2}-5\,\hat{I}_{{0}}(3,1,1){m_{c}}\,{{m_{b}}}^{2}+20\,\hat{I}_{{0}}(1,3,1){{m_{b}%
}}^{3} \\
&&-5\,\hat{I}_{{0}}(2,2,1){{m_{b}}}^{3}-20\,\hat{I}_{{6}}(3,2,1){{m_{b}}}%
^{3}-15\,\hat{I}_{0}^{[0,1]}(3,2,1){{m_{b}}}^{3}+30\,\hat{I}_{0}^{[0,1]}(1,4,1){{m_{b}}}%
^{3} \\
&&+5\,\hat{I}_{0}^{[0,2]}(3,2,2){{m_{b}}}^{3}+5\,\hat{I}_{{0}}(1,2,2){{m_{b}}}%
^{3}+120\,\hat{I}_{{6}}(1,4,1){{m_{b}}}^{3}-10\,\hat{I}_{0}^{[0,1]}(2,3,1){{m_{b}}}^{3}
\\
&&-20\,\hat{I}_{{6}}(2,2,2){{m_{b}}}^{3}+20\,\hat{I}_{6}^{[0,1]}(3,2,2){{m_{b}}}%
^{3}-40\,\hat{I}_{{6}}(2,3,1){{m_{b}}}^{3}-10\,\hat{I}_{0}^{[0,1]}(2,2,2){{m_{b}}}^{3} \\
&&-5\,\hat{I}_{{0}}(1,1,2){m_{c}}+20\,\hat{I}_{6}^{[0,2]}(3,2,2){m_{c}}-40\,\hat{I}_{{6}%
}(3,1,1){m_{c}}-20\,\hat{I}_{{6}}(2,2,1){m_{c}} \\
&&-40\,\hat{I}_{6}^{[0,1]}(2,2,2){m_{c}}-20\,\hat{I}_{6}^{[0,1]}(3,1,2){m_{c}}-5\,\hat{I}_{{0}%
}(1,2,1){m_{c}}-5\,\hat{I}_{0}^{[0,1]}(3,1,1){m_{c}} \\
&&-40\,\hat{I}_{6}^{[0,1]}(3,2,1){m_{c}}+5\,\hat{I}_{{0}}(2,1,1){m_{c}}-20\,\hat{I}_{{6}%
}(2,1,2){m_{c}}+40\,\hat{I}_{{6}}(1,3,1){m_{b}} \\
&&-15\,\hat{I}_{0}^{[0,1]}(2,2,1){m_{b}}+15\,\hat{I}_{0}^{[0,2]}(2,2,2){m_{b}}+15\,\hat{I}_{{0}%
}(1,1,2){m_{b}}+40\,\hat{I}_{6}^{[0,1]}(2,3,1){m_{b}} \\
&&+20\,\hat{I}_{6}^{[0,1]}(3,2,1){m_{b}}+20\,\hat{I}_{{6}}(2,1,2){m_{b}}%
+10\,\hat{I}_{0}^{[0,2]}(2,3,1){m_{b}}+15\,\hat{I}_{0}^{[0,2]}(3,2,1){m_{b}} \\
&&+40\,\hat{I}_{{6}}(2,2,1){m_{b}}-20\,\hat{I}_{0}^{[0,1]}(1,2,2){m_{b}}%
+10\,\hat{I}_{0}^{[0,2]}(3,1,2){m_{b}}+20\,\hat{I}_{{6}}(1,2,2){m_{b}} \\
&&+20\,\hat{I}_{6}^{[0,1]}(2,2,2){m_{b}}+40\,\hat{I}_{{6}}(3,1,1){m_{b}}%
+40\,\hat{I}_{6}^{[0,1]}(3,1,2){m_{b}}+10\,\hat{I}_{0}^{[0,1]}(1,3,1){m_{b}} \\
&&+15\,\hat{I}_{{0}}(1,2,1){m_{b}}-15\,\hat{I}_{{0}}(2,1,1){m_{b}}-10%
\,\hat{I}_{0}^{[0,1]}(3,1,1){m_{b}}-20\,\hat{I}_{6}^{[0,2]}(3,2,2){m_{b}}
\end{eqnarray*}
\begin{eqnarray*}
C_4^1&=&5\,\hat{I}_{{1}}(3,2,2){{m_{c}}}^{5}+10\,\hat{I}_{{3}}(3,2,2){{m_{c}}}^{5}+10\,\hat{I}_{{%
4}}(3,2,2){{m_{c}}}^{5}+5\,\hat{I}_{{2}}(3,2,2){{m_{c}}}^{5} \\
&&-10\,\hat{I}_{{4}}(3,2,2){{m_{c}}}^{4}{m_{b}}-10\,\hat{I}_{{3}}(3,2,2){{m_{c}}}^{4}{%
m_{b}}-10\,\hat{I}_{{3}}(3,2,2){{m_{c}}}^{3}{{m_{b}}}^{2}-10\,\hat{I}_{{4}}(3,2,2){{m_{c}%
}}^{3}{{m_{b}}}^{2} \\
&&-5\,\hat{I}_{{1}}(3,2,2){{m_{c}}}^{3}{{m_{b}}}^{2}-5\,\hat{I}_{{2}}(3,2,2){{m_{c}}}^{3}%
{{m_{b}}}^{2}+10\,\hat{I}_{{4}}(3,2,2){{m_{c}}}^{2}{{m_{b}}}^{3}+10\,\hat{I}_{{3}}(3,2,2)%
{{m_{c}}}^{2}{{m_{b}}}^{3} \\
&&+15\,\hat{I}_{{2}}(4,1,1){{m_{c}}}^{3}+20\,\hat{I}_{{4}}(3,2,1){{m_{c}}}^{3}+15\,\hat{I}_{{1}%
}(4,1,1){{m_{c}}}^{3}+20\,\hat{I}_{{2}}(3,2,1){{m_{c}}}^{3} \\
&&+20\,\hat{I}_{{3}}(2,2,2){{m_{c}}}^{3}+30\,\hat{I}_{{3}}(4,1,1){{m_{c}}}^{3}+20\,\hat{I}_{{1}%
}(3,2,1){{m_{c}}}^{3}+10\,\hat{I}_{{2}}(2,2,2){{m_{c}}}^{3} \\
&&-10\,\hat{I}_{6}^{[0,1]}(3,2,2){{m_{c}}}^{3}-10\,\hat{I}_{1}^{[0,1]}(3,2,2){{m_{c}}}%
^{3}+20\,\hat{I}_{{4}}(2,2,2){{m_{c}}}^{3}+10\,\hat{I}_{{4}}(3,1,2){{m_{c}}}^{3} \\
&&+10\,\hat{I}_{{3}}(3,1,2){{m_{c}}}^{3}+5\,\hat{I}_{{1}}(3,1,2){{m_{c}}}^{3}+20\,\hat{I}_{{3}%
}(3,2,1){{m_{c}}}^{3}-20\,\hat{I}_{3}^{[0,1]}(3,2,2){{m_{c}}}^{3} \\
&&-20\,\hat{I}_{4}^{[0,1]}(3,2,2){{m_{c}}}^{3}+5\,\hat{I}_{{2}}(3,1,2){{m_{c}}}%
^{3}+10\,\hat{I}_{{1}}(2,2,2){{m_{c}}}^{3}+20\,\hat{I}_{{0}}(3,2,1){{m_{c}}}^{3} \\
&&+30\,\hat{I}_{{4}}(4,1,1){{m_{c}}}^{3}-10\,\hat{I}_{{3}}(2,2,2){{m_{c}}}^{2}{m_{b}}%
-20\,\hat{I}_{{4}}(3,2,1){{m_{c}}}^{2}{m_{b}}-20\,\hat{I}_{{3}}(3,2,1){{m_{c}}}^{2}{m_{b}%
} \\
&&-15\,\hat{I}_{{2}}(3,2,1){{m_{c}}}^{2}{m_{b}}+20\,\hat{I}_{{1}}(2,3,1){{m_{c}}}^{2}{%
m_{b}}+20\,\hat{I}_{4}^{[0,1]}(3,2,2){{m_{c}}}^{2}{m_{b}}+20\,\hat{I}_{3}^{[0,1]}(3,2,2){%
{m_{c}}}^{2}{m_{b}} \\
&&-10\,\hat{I}_{{3}}(3,1,2){{m_{c}}}^{2}{m_{b}}-10\,\hat{I}_{{4}}(3,1,2){{m_{c}}}^{2}{%
m_{b}}-10\,\hat{I}_{{4}}(2,2,2){{m_{c}}}^{2}{m_{b}}-30\,\hat{I}_{{4}}(4,1,1){{m_{c}}}^{2}%
{m_{b}} \\
&&+20\,\hat{I}_{{2}}(2,3,1){{m_{c}}}^{2}{m_{b}}-30\,\hat{I}_{{3}}(4,1,1){{m_{c}}}^{2}{%
m_{b}}+20\,\hat{I}_{{3}}(2,3,1){{m_{c}}}^{2}{m_{b}}-15\,\hat{I}_{{1}}(3,2,1){{m_{c}}}^{2}%
{m_{b}} \\
&&+20\,\hat{I}_{{4}}(2,3,1){{m_{c}}}^{2}{m_{b}}+10\,\hat{I}_{{1}}(3,2,1){m_{c}}\,{{m_{b}}%
}^{2}-60\,\hat{I}_{{4}}(1,4,1){m_{c}}\,{{m_{b}}}^{2}-30\,\hat{I}_{{1}}(1,4,1){m_{c}}\,{{%
m_{b}}}^{2} \\
&&+10\,\hat{I}_{{4}}(3,2,1){m_{c}}\,{{m_{b}}}^{2}+10\,\hat{I}_{{3}}(2,2,2){m_{c}}\,{{%
m_{b}}}^{2}-10\,\hat{I}_{3}^{[0,1]}(3,2,2){m_{c}}\,{{m_{b}}}^{2}-60\,\hat{I}_{{3}}(1,4,1)%
{m_{c}}\,{{m_{b}}}^{2} \\
&&+10\,\hat{I}_{{3}}(3,2,1){m_{c}}\,{{m_{b}}}^{2}-5\,\hat{I}_{6}^{[0,1]}(3,2,2){m_{c}}\,{%
{m_{b}}}^{2}-10\,\hat{I}_{4}^{[0,1]}(3,2,2){m_{c}}\,{{m_{b}}}^{2}-5%
\,\hat{I}_{1}^{[0,1]}(3,2,2){m_{c}}\,{{m_{b}}}^{2} \\
&&-30\,\hat{I}_{{2}}(1,4,1){m_{c}}\,{{m_{b}}}^{2}+10\,\hat{I}_{{4}}(2,2,2){m_{c}}\,{{%
m_{b}}}^{2}+10\,\hat{I}_{{2}}(3,2,1){m_{c}}\,{{m_{b}}}^{2}-10\,\hat{I}_{{3}}(2,2,2){{%
m_{b}}}^{3} \\
&&-10\,\hat{I}_{{3}}(3,2,1){{m_{b}}}^{3}-20\,\hat{I}_{{4}}(2,3,1){{m_{b}}}%
^{3}+10\,\hat{I}_{4}^{[0,1]}(3,2,2){{m_{b}}}^{3}-20\,\hat{I}_{{3}}(2,3,1){{m_{b}}}^{3} \\
&&-10\,\hat{I}_{{4}}(2,2,2){{m_{b}}}^{3}+60\,\hat{I}_{{3}}(1,4,1){{m_{b}}}%
^{3}+10\,\hat{I}_{3}^{[0,1]}(3,2,2){{m_{b}}}^{3}-10\,\hat{I}_{{4}}(3,2,1){{m_{b}}}^{3} \\
&&+60\,\hat{I}_{{4}}(1,4,1){{m_{b}}}^{3}-20\,\hat{I}_{{3}}(3,1,1){m_{c}}%
-10\,\hat{I}_{1}^{[0,1]}(3,2,1){m_{c}}+5\,\hat{I}_{2}^{[0,2]}(3,2,2){m_{c}} \\
&&-10\,\hat{I}_{{4}}(2,1,2){m_{c}}+10\,\hat{I}_{3}^{[0,2]}(3,2,2){m_{c}}-5\,\hat{I}_{{2}%
}(3,1,1){m_{c}}-10\,\hat{I}_{3}^{[0,1]}(3,1,2){m_{c}} \\
&&-10\,\hat{I}_{6}^{[0,1]}(2,2,2){m_{c}}-10\,\hat{I}_{{3}}(2,1,2){m_{c}}%
-20\,\hat{I}_{3}^{[0,1]}(2,2,2){m_{c}}-10\,\hat{I}_{{3}}(2,2,1){m_{c}} \\
&&-10\,\hat{I}_{1}^{[0,1]}(2,2,2){m_{c}}+10\,\hat{I}_{4}^{[0,2]}(3,2,2){m_{c}}%
-20\,\hat{I}_{3}^{[0,1]}(3,2,1){m_{c}}-10\,\hat{I}_{{4}}(2,2,1){m_{c}} \\
&&+15\,\hat{I}_{{1}}(2,1,2){m_{c}}-10\,\hat{I}_{6}^{[0,1]}(3,2,1){m_{c}}%
+5\,\hat{I}_{1}^{[0,2]}(3,2,2){m_{c}}-20\,\hat{I}_{4}^{[0,1]}(2,2,2){m_{c}} \\
&&+10\,\hat{I}_{{0}}(2,2,1){m_{c}}+15\,\hat{I}_{{2}}(2,1,2){m_{c}}-20\,\hat{I}_{{4}}(3,1,1){%
m_{c}}-15\,\hat{I}_{1}^{[0,1]}(3,1,2){m_{c}} \\
&&-15\,\hat{I}_{6}^{[0,1]}(3,1,2){m_{c}}-20\,\hat{I}_{4}^{[0,1]}(3,2,1){m_{c}}%
-10\,\hat{I}_{4}^{[0,1]}(3,1,2){m_{c}}-5\,\hat{I}_{{1}}(3,1,1){m_{c}} \\
&&+20\,\hat{I}_{4}^{[0,1]}(2,3,1){m_{b}}+20\,\hat{I}_{{3}}(3,1,1){m_{b}}+10\,\hat{I}_{{4}%
}(2,1,2){m_{b}}-10\,\hat{I}_{4}^{[0,2]}(3,2,2){m_{b}} \\
&&+20\,\hat{I}_{{3}}(1,3,1){m_{b}}+10\,\hat{I}_{3}^{[0,1]}(3,2,1){m_{b}}+10\,\hat{I}_{{3}%
}(2,1,2){m_{b}}+10\,\hat{I}_{4}^{[0,1]}(3,2,1){m_{b}} \\
&&-10\,\hat{I}_{3}^{[0,2]}(3,2,2){m_{b}}-10\,\hat{I}_{{2}}(1,3,1){m_{b}}%
+20\,\hat{I}_{3}^{[0,1]}(2,3,1){m_{b}}+10\,\hat{I}_{{1}}(2,2,1){m_{b}} \\
&&+10\,\hat{I}_{3}^{[0,1]}(2,2,2){m_{b}}+20\,\hat{I}_{{4}}(3,1,1){m_{b}}+20\,\hat{I}_{{3}%
}(2,2,1){m_{b}}+20\,\hat{I}_{{4}}(2,2,1){m_{b}} \\
&&+10\,\hat{I}_{{2}}(2,2,1){m_{b}}+10\,\hat{I}_{{3}}(1,2,2){m_{b}}+10\,\hat{I}_{{4}}(1,2,2){%
m_{b}}+20\,\hat{I}_{{4}}(1,3,1){m_{b}}
\end{eqnarray*}
\begin{eqnarray*}
C_4^2&=&5\,\hat{I}_{{1}}(3,2,2){{m_{c}}}^{5}+10\,\hat{I}_{{3}}(3,2,2){{m_{c}}}^{5}-5\,\hat{I}_{{2%
}}(3,2,2){{m_{c}}}^{5}-10\,\hat{I}_{{4}}(3,2,2){{m_{c}}}^{5} \\
&&+10\,\hat{I}_{{4}}(3,2,2){{m_{c}}}^{4}{m_{b}}-10\,\hat{I}_{{3}}(3,2,2){{m_{c}}}^{4}{%
m_{b}}-5\,\hat{I}_{{1}}(3,2,2){{m_{c}}}^{3}{{m_{b}}}^{2}+10\,\hat{I}_{{4}}(3,2,2){{m_{c}}%
}^{3}{{m_{b}}}^{2} \\
&&-10\,\hat{I}_{{3}}(3,2,2){{m_{c}}}^{3}{{m_{b}}}^{2}+5\,\hat{I}_{{2}}(3,2,2){{m_{c}}}%
^{3}{{m_{b}}}^{2}-10\,\hat{I}_{{4}}(3,2,2){{m_{c}}}^{2}{{m_{b}}}^{3}+10\,\hat{I}_{{3}%
}(3,2,2){{m_{c}}}^{2}{{m_{b}}}^{3} \\
&&+15\,\hat{I}_{{1}}(4,1,1){{m_{c}}}^{3}+5\,\hat{I}_{{1}}(3,1,2){{m_{c}}}%
^{3}-10\,\hat{I}_{1}^{[0,1]}(3,2,2){{m_{c}}}^{3}+20\,\hat{I}_{{3}}(2,2,2){{m_{c}}}^{3} \\
&&-30\,\hat{I}_{{4}}(4,1,1){{m_{c}}}^{3}+20\,\hat{I}_{{0}}(3,2,1){{m_{c}}}^{3}-20\,\hat{I}_{{2}%
}(3,2,1){{m_{c}}}^{3}+20\,\hat{I}_{{3}}(3,2,1){{m_{c}}}^{3} \\
&&+20\,\hat{I}_{{1}}(3,2,1){{m_{c}}}^{3}+10\,\hat{I}_{2}^{[0,1]}(3,2,2){{m_{c}}}%
^{3}-20\,\hat{I}_{{4}}(3,2,1){{m_{c}}}^{3}-20\,\hat{I}_{{4}}(2,2,2){{m_{c}}}^{3} \\
&&+30\,\hat{I}_{{3}}(4,1,1){{m_{c}}}^{3}+10\,\hat{I}_{{1}}(2,2,2){{m_{c}}}^{3}-15\,\hat{I}_{{2}%
}(4,1,1){{m_{c}}}^{3}+20\,\hat{I}_{4}^{[0,1]}(3,2,2){{m_{c}}}^{3} \\
&&-5\,\hat{I}_{{2}}(3,1,2){{m_{c}}}^{3}+10\,\hat{I}_{{3}}(3,1,2){{m_{c}}}^{3}-10\,\hat{I}_{{2}%
}(2,2,2){{m_{c}}}^{3}-10\,\hat{I}_{{4}}(3,1,2){{m_{c}}}^{3} \\
&&-20\,\hat{I}_{3}^{[0,1]}(3,2,2){{m_{c}}}^{3}-15\,\hat{I}_{{1}}(3,2,1){{m_{c}}}^{2}{%
m_{b}}-10\,\hat{I}_{{3}}(3,1,2){{m_{c}}}^{2}{m_{b}}+20\,\hat{I}_{{4}}(3,2,1){{m_{c}}}^{2}%
{m_{b}} \\
&&-20\,\hat{I}_{{3}}(3,2,1){{m_{c}}}^{2}{m_{b}}-20\,\hat{I}_{{4}}(2,3,1){{m_{c}}}^{2}{%
m_{b}}+20\,\hat{I}_{{3}}(2,3,1){{m_{c}}}^{2}{m_{b}}-20\,\hat{I}_{4}^{[0,1]}(3,2,2){{m_{c}%
}}^{2}{m_{b}} \\
&&-20\,\hat{I}_{{2}}(2,3,1){{m_{c}}}^{2}{m_{b}}-10\,\hat{I}_{{3}}(2,2,2){{m_{c}}}^{2}{%
m_{b}}+15\,\hat{I}_{{2}}(3,2,1){{m_{c}}}^{2}{m_{b}}+20\,\hat{I}_{3}^{[0,1]}(3,2,2){{m_{c}%
}}^{2}{m_{b}} \\
&&+10\,\hat{I}_{{4}}(2,2,2){{m_{c}}}^{2}{m_{b}}+30\,\hat{I}_{{4}}(4,1,1){{m_{c}}}^{2}{%
m_{b}}+20\,\hat{I}_{{1}}(2,3,1){{m_{c}}}^{2}{m_{b}}-30\,\hat{I}_{{3}}(4,1,1){{m_{c}}}^{2}%
{m_{b}} \\
&&+10\,\hat{I}_{{4}}(3,1,2){{m_{c}}}^{2}{m_{b}}+30\,\hat{I}_{{2}}(1,4,1){m_{c}}\,{{m_{b}}%
}^{2}-5\,\hat{I}_{1}^{[0,1]}(3,2,2){m_{c}}\,{{m_{b}}}^{2}+10\,\hat{I}_{{1}}(3,2,1){m_{c}}%
\,{{m_{b}}}^{2} \\
&&+60\,\hat{I}_{{4}}(1,4,1){m_{c}}\,{{m_{b}}}^{2}-60\,\hat{I}_{{3}}(1,4,1){m_{c}}\,{{%
m_{b}}}^{2}+5\,\hat{I}_{2}^{[0,1]}(3,2,2){m_{c}}\,{{m_{b}}}^{2}-10\,\hat{I}_{{4}}(3,2,1){%
m_{c}}\,{{m_{b}}}^{2} \\
&&-10\,\hat{I}_{{4}}(2,2,2){m_{c}}\,{{m_{b}}}^{2}-30\,\hat{I}_{{1}}(1,4,1){m_{c}}\,{{%
m_{b}}}^{2}+10\,\hat{I}_{{3}}(3,2,1){m_{c}}\,{{m_{b}}}^{2}+10\,\hat{I}_{{3}}(2,2,2){m_{c}%
}\,{{m_{b}}}^{2} \\
&&+10\,\hat{I}_{4}^{[0,1]}(3,2,2){m_{c}}\,{{m_{b}}}^{2}-10\,\hat{I}_{{2}}(3,2,1){m_{c}}\,%
{{m_{b}}}^{2}-10\,\hat{I}_{3}^{[0,1]}(3,2,2){m_{c}}\,{{m_{b}}}^{2}-10%
\,\hat{I}_{4}^{[0,1]}(3,2,2){{m_{b}}}^{3} \\
&&-60\,\hat{I}_{{4}}(1,4,1){{m_{b}}}^{3}-10\,\hat{I}_{{3}}(3,2,1){{m_{b}}}^{3}+10\,\hat{I}_{{4}%
}(3,2,1){{m_{b}}}^{3}-20\,\hat{I}_{{3}}(2,3,1){{m_{b}}}^{3} \\
&&+60\,\hat{I}_{{3}}(1,4,1){{m_{b}}}^{3}-10\,\hat{I}_{{3}}(2,2,2){{m_{b}}}^{3}+10\,\hat{I}_{{4}%
}(2,2,2){{m_{b}}}^{3}+10\,\hat{I}_{3}^{[0,1]}(3,2,2){{m_{b}}}^{3} \\
&&+20\,\hat{I}_{{4}}(2,3,1){{m_{b}}}^{3}+20\,\hat{I}_{{4}}(3,1,1){m_{c}}%
+5\,\hat{I}_{1}^{[0,2]}(3,2,2){m_{c}}-10\,\hat{I}_{1}^{[0,1]}(2,2,2){m_{c}} \\
&&+10\,\hat{I}_{2}^{[0,1]}(2,2,2){m_{c}}+20\,\hat{I}_{4}^{[0,1]}(3,2,1){m_{c}}%
+10\,\hat{I}_{4}^{[0,1]}(3,1,2){m_{c}}-10\,\hat{I}_{1}^{[0,1]}(3,2,1){m_{c}} \\
&&+10\,\hat{I}_{2}^{[0,1]}(3,2,1){m_{c}}-10\,\hat{I}_{4}^{[0,2]}(3,2,2){m_{c}}-10\,\hat{I}_{{3}%
}(2,1,2){m_{c}}+10\,\hat{I}_{{4}}(2,2,1){m_{c}} \\
&&+15\,\hat{I}_{2}^{[0,1]}(3,1,2){m_{c}}-15\,\hat{I}_{{2}}(2,1,2){m_{c}}+10\,\hat{I}_{{4}%
}(2,1,2){m_{c}}+5\,\hat{I}_{{2}}(3,1,1){m_{c}} \\
&&+15\,\hat{I}_{{1}}(2,1,2){m_{c}}-5\,\hat{I}_{2}^{[0,2]}(3,2,2){m_{c}}%
-15\,\hat{I}_{1}^{[0,1]}(3,1,2){m_{c}}+10\,\hat{I}_{{0}}(2,2,1){m_{c}} \\
&&+20\,\hat{I}_{4}^{[0,1]}(2,2,2){m_{c}}-20\,\hat{I}_{3}^{[0,1]}(2,2,2){m_{c}}%
-10\,\hat{I}_{3}^{[0,1]}(3,1,2){m_{c}}-20\,\hat{I}_{3}^{[0,1]}(3,2,1){m_{c}} \\
&&+10\,\hat{I}_{3}^{[0,2]}(3,2,2){m_{c}}-5\,\hat{I}_{{1}}(3,1,1){m_{c}}-10\,\hat{I}_{{3}%
}(2,2,1){m_{c}}-20\,\hat{I}_{{3}}(3,1,1){m_{c}} \\
&&+10\,\hat{I}_{{1}}(2,2,1){m_{b}}+10\,\hat{I}_{{3}}(2,1,2){m_{b}}+10%
\,\hat{I}_{4}^{[0,2]}(3,2,2){m_{b}}-10\,\hat{I}_{{2}}(2,2,1){m_{b}} \\
&&-10\,\hat{I}_{4}^{[0,1]}(2,2,2){m_{b}}+10\,\hat{I}_{{2}}(1,3,1){m_{b}}%
+10\,\hat{I}_{3}^{[0,1]}(2,2,2){m_{b}}-20\,\hat{I}_{{4}}(3,1,1){m_{b}} \\
&&-10\,\hat{I}_{{1}}(1,3,1){m_{b}}+10\,\hat{I}_{{3}}(1,2,2){m_{b}}-10\,\hat{I}_{{4}}(1,2,2){%
m_{b}}-20\,\hat{I}_{4}^{[0,1]}(3,1,2){m_{b}} \\
&&-10\,\hat{I}_{4}^{[0,1]}(3,2,1){m_{b}}-10\,\hat{I}_{3}^{[0,2]}(3,2,2){m_{b}}%
+10\,\hat{I}_{3}^{[0,1]}(3,2,1){m_{b}}-20\,\hat{I}_{{4}}(2,2,1){m_{b}} \\
&&+20\,\hat{I}_{{3}}(1,3,1){m_{b}}-20\,\hat{I}_{{4}}(1,3,1){m_{b}}-10\,\hat{I}_{{4}}(2,1,2){%
m_{b}}+20\,\hat{I}_{{3}}(3,1,1){m_{b}}
\end{eqnarray*}
where
\begin{eqnarray}
\hat{I}_n^{[i,j]} (a,b,c) = \left( M_1^2 \right)^i \left( M_2^2
\right)^j
\frac{d^i}{d\left( M_1^2 \right)^i} \frac{d^j}{d\left( M_2^2 \right)^j} %
\left[\left( M_1^2 \right)^i \left( M_2^2 \right)^j \hat{I}_n(a,b,c) \right]%
~.  \nonumber
\end{eqnarray}
\newpage
\appendix

\begin{center}
{\Large \textbf{Appendix--B}}
\end{center}

\setcounter{equation}{0} \renewcommand{\theequation}

In this appendix, the explicit expressions of the coefficients of
the gluon condensate  entering  the HQET limit of the  form
factors $f_{V}^{HQET}, f_{0}^{HQET}, f_{1}^{HQET}$ and
$f_{2}^{HQET}$ are presented.

\begin{eqnarray*}
C^{HQET}_{V}=&&60\,{\frac{\bar{I}_{{1}}(1,4,1){{m_{b}}}^{3}}{\sqrt{Z}}}-30\,{\frac{\bar{I}%
_{{1}}(3,2,1){{m_{b}}}^{3}}{\sqrt{Z}}}-20\,{\frac{\bar{I}_{{2}}(3,2,1){{m_{b}%
}}^{3}}{\sqrt{Z}}} \\
&&+10\,{\frac{\bar{I}_{0}^{[0,1]}(3,2,2){{m_{b}}}^{3}}{\sqrt{Z}}}+60\,{\frac{%
\bar{I}_{{0}}(1,4,1){{m_{b}}}^{3}}{\sqrt{Z}}}+10\,{\frac{\bar{I}%
_{2}^{[0,1]}(3,2,2){{m_{b}}}^{3}}{\sqrt{Z}}} \\
&&+10\,{\frac{\bar{I}_{1}^{[0,1]}(3,2,2){{m_{b}}}^{3}}{\sqrt{Z}}}-20\,{\frac{%
\bar{I}_{{0}}(3,2,1){{m_{b}}}^{3}}{\sqrt{Z}}}-20\,{\frac{\bar{I}%
_{1}^{[0,1]}(3,2,2){{m_{b}}}^{3}}{Z}} \\
&&+30\,{\frac{\bar{I}_{{1}}(4,1,1){{m_{b}}}^{3}}{Z}}-10\,{\frac{\bar{I}_{{1}%
}(3,2,2){{m_{b}}}^{5}}{Z}}+40\,{\frac{\bar{I}_{{0}}(2,3,1){{m_{b}}}^{3}}{Z}}
\\
&&+20\,{\frac{\bar{I}_{{1}}(2,2,2){{m_{b}}}^{3}}{Z}}+10\,{\frac{\bar{I}_{{0}%
}(3,2,1){{m_{b}}}^{3}}{Z}}+10\,{\frac{\bar{I}_{{1}}(3,2,1){{m_{b}}}^{3}}{Z}}
\\
&&-10\,{\frac{\bar{I}_{{2}}(3,2,1){{m_{b}}}^{3}}{Z}}+20\,{\frac{\bar{I}_{{1}%
}(2,3,1){{m_{b}}}^{3}}{Z}}-30\,{\frac{\bar{I}_{{1}}(4,1,1){{m_{b}}}^{3}}{{Z}%
^{3/2}}} \\
&&+10\,{\frac{\bar{I}_{{0}}(3,2,2){{m_{b}}}^{5}}{{Z}^{3/2}}}+10\,{\frac{%
\bar{I}_{{2}}(3,2,2){{m_{b}}}^{5}}{{Z}^{3/2}}}-20\,{\frac{\bar{I}_{{2}%
}(2,2,2){{m_{b}}}^{3}}{{Z}^{3/2}}} \\
&&-30\,{\frac{\bar{I}_{{2}}(4,1,1){{m_{b}}}^{3}}{{Z}^{3/2}}}+10\,{\frac{%
\bar{I}_{{1}}(3,2,2){{m_{b}}}^{5}}{{Z}^{3/2}}}-20\,{\frac{\bar{I}_{{1}%
}(2,2,2){{m_{b}}}^{3}}{{Z}^{3/2}}} \\
&&-10\,{\frac{\bar{I}_{{0}}(3,1,2){{m_{b}}}^{3}}{{Z}^{3/2}}}+20\,{\frac{%
\bar{I}_{2}^{[0,1]}(3,2,2){{m_{b}}}^{3}}{{Z}^{3/2}}}-20\,{\frac{\bar{I}_{{0}%
}(2,2,2){{m_{b}}}^{3}}{{Z}^{3/2}}} \\
&&-30\,{\frac{\bar{I}_{{0}}(4,1,1){{m_{b}}}^{3}}{{Z}^{3/2}}}+20\,{\frac{%
\bar{I}_{0}^{[0,1]}(3,2,2){{m_{b}}}^{3}}{{Z}^{3/2}}}+10\,{\frac{\bar{I}_{{2}%
}(3,2,1){{m_{b}}}^{3}}{{Z}^{3/2}}} \\
&&-10\,{\frac{\bar{I}_{{2}}(3,1,2){{m_{b}}}^{3}}{{Z}^{3/2}}}+20\,{\frac{%
\bar{I}_{1}^{[0,1]}(3,2,2){{m_{b}}}^{3}}{{Z}^{3/2}}}+10\,{\frac{\bar{I}_{{1}%
}(3,2,2){{m_{b}}}^{5}}{{Z}^{2}}} \\
&&-10\,{\frac{\bar{I}_{{1}}(3,2,2){{m_{b}}}^{5}}{{Z}^{5/2}}}-10\,{\frac{%
\bar{I}_{{0}}(3,2,2){{m_{b}}}^{5}}{{Z}^{5/2}}}-10\,{\frac{\bar{I}_{{2}%
}(3,2,2){{m_{b}}}^{5}}{{Z}^{5/2}}} \\
&&+10\,\bar{I}_{{1}}(2,2,2){{m_{b}}}^{3}+20\,\bar{I}_{{1}}(3,2,1){{m_{b}}}%
^{3}-60\,\bar{I}_{{1}}(1,4,1){{m_{b}}}^{3} \\
&&-20\,\bar{I}_{{1}}(2,3,1){{m_{b}}}^{3}-10\,\bar{I}_{1}^{[0,1]}(3,2,2){{%
m_{b}}}^{3}
\end{eqnarray*}
\begin{eqnarray*}
C^{HQET}_{0}=&&-10\,{\frac{\bar{I}_{{0}}(3,2,1){{m_{b}}}^{5}}{\sqrt{Z}}}+5\,{\frac{\bar{I}%
_{0}^{[0,1]}(3,2,2){{m_{b}}}^{5}}{\sqrt{Z}}}+30\,{\frac{\bar{I}_{{0}}(1,4,1){%
{m_{b}}}^{5}}{\sqrt{Z}}} \\
&&-10\,{\frac{\bar{I}_{{0}}(2,3,1){{m_{b}}}^{5}}{Z}}+20\,{\frac{\bar{I}_{{6}%
}(3,2,2){{m_{b}}}^{5}}{Z}}-30\,{\frac{\bar{I}_{{0}}(1,4,1){{m_{b}}}^{5}}{Z}}
\\
&&+10\,{\frac{\bar{I}_{{0}}(3,2,1){{m_{b}}}^{5}}{Z}}+10\,{\frac{\bar{I}%
_{0}^{[0,1]}(3,2,2){{m_{b}}}^{5}}{{Z}^{3/2}}}-5\,{\frac{\bar{I}_{{0}}(3,2,1){%
{m_{b}}}^{5}}{{Z}^{3/2}}} \\
&&-20\,{\frac{\bar{I}_{{6}}(3,2,2){{m_{b}}}^{5}}{{Z}^{3/2}}}-15\,{\frac{%
\bar{I}_{{0}}(4,1,1){{m_{b}}}^{5}}{{Z}^{3/2}}}-10\,{\frac{\bar{I}_{{0}%
}(2,2,2){{m_{b}}}^{5}}{{Z}^{3/2}}} \\
&&+5\,{\frac{\bar{I}_{{0}}(3,2,2){{m_{b}}}^{7}}{{Z}^{3/2}}}+15\,{\frac{%
\bar{I}_{{0}}(4,1,1){{m_{b}}}^{5}}{{Z}^{2}}}+15\,{\frac{\bar{I}_{{0}}(2,2,2){%
{m_{b}}}^{5}}{{Z}^{2}}} \\
&&-15\,{\frac{\bar{I}_{0}^{[0,1]}(3,2,2){{m_{b}}}^{5}}{{Z}^{2}}}+10\,{\frac{%
\bar{I}_{{0}}(2,3,1){{m_{b}}}^{5}}{{Z}^{2}}}-5\,{\frac{\bar{I}_{{0}}(3,2,2){{%
m_{b}}}^{7}}{{Z}^{2}}} \\
&&-20\,{\frac{\bar{I}_{{6}}(3,2,2){{m_{b}}}^{5}}{{Z}^{2}}}+5\,{\frac{\bar{I}%
_{{0}}(3,2,1){{m_{b}}}^{5}}{{Z}^{2}}}-5\,{\frac{\bar{I}_{{0}}(3,2,2){{m_{b}}}%
^{7}}{{Z}^{5/2}}} \\
&&+20\,{\frac{\bar{I}_{{6}}(3,2,2){{m_{b}}}^{5}}{{Z}^{5/2}}}+5\,{\frac{%
\bar{I}_{{0}}(3,2,2){{m_{b}}}^{7}}{{Z}^{3}}}
\end{eqnarray*}
\begin{eqnarray*}
C^{HQET}_{1}=&&40\,{\frac{\bar{I}_{{1}}(3,2,1){{m_{b}}}^{3}}{\sqrt{Z}}}+10\,{\frac{\bar{I}%
_{{1}}(2,2,2){{m_{b}}}^{3}}{\sqrt{Z}}}-5\,{\frac{\bar{I}_{2}^{[0,1]}(3,2,2){{%
m_{b}}}^{3}}{\sqrt{Z}}} \\
&&-20\,{\frac{\bar{I}_{1}^{[0,1]}(3,2,2){{m_{b}}}^{3}}{\sqrt{Z}}}+20\,{\frac{%
\bar{I}_{{4}}(3,2,1){{m_{b}}}^{3}}{\sqrt{Z}}}-20\,{\frac{\bar{I}%
_{4}^{[0,1]}(3,2,2){{m_{b}}}^{3}}{\sqrt{Z}}} \\
&&+15\,{\frac{\bar{I}_{{0}}(3,2,1){{m_{b}}}^{3}}{\sqrt{Z}}}+20\,{\frac{%
\bar{I}_{{4}}(2,2,2){{m_{b}}}^{3}}{\sqrt{Z}}}+20\,{\frac{\bar{I}_{{3}}(2,2,2)%
{{m_{b}}}^{3}}{\sqrt{Z}}} \\
&&-5\,{\frac{\bar{I}_{{1}}(3,1,2){{m_{b}}}^{3}}{\sqrt{Z}}}+10\,{\frac{\bar{I}%
_{{2}}(3,2,1){{m_{b}}}^{3}}{\sqrt{Z}}}-20\,{\frac{\bar{I}_{3}^{[0,1]}(3,2,2){%
{m_{b}}}^{3}}{\sqrt{Z}}} \\
&&-120\,{\frac{\bar{I}_{{1}}(1,4,1){{m_{b}}}^{3}}{\sqrt{Z}}}-120\,{\frac{%
\bar{I}_{{3}}(1,4,1){{m_{b}}}^{3}}{\sqrt{Z}}}-30\,{\frac{\bar{I}_{{0}}(1,4,1)%
{{m_{b}}}^{3}}{\sqrt{Z}}} \\
&&+20\,{\frac{\bar{I}_{{3}}(3,2,1){{m_{b}}}^{3}}{\sqrt{Z}}}-120\,{\frac{%
\bar{I}_{{4}}(1,4,1){{m_{b}}}^{3}}{\sqrt{Z}}}-5\,{\frac{\bar{I}%
_{0}^{[0,1]}(3,2,2){{m_{b}}}^{3}}{\sqrt{Z}}} \\
&&-30\,{\frac{\bar{I}_{{2}}(1,4,1){{m_{b}}}^{3}}{\sqrt{Z}}}+40\,{\frac{%
\bar{I}_{{4}}(2,3,1){{m_{b}}}^{3}}{Z}}+40\,{\frac{\bar{I}_{{3}}(2,3,1){{m_{b}%
}}^{3}}{Z}} \\
&&+70\,{\frac{\bar{I}_{{1}}(2,3,1){{m_{b}}}^{3}}{Z}}-50\,{\frac{\bar{I}_{{1}%
}(3,2,1){{m_{b}}}^{3}}{Z}}-20\,{\frac{\bar{I}_{{4}}(3,1,2){{m_{b}}}^{3}}{Z}}
\\
&&-10\,{\frac{\bar{I}_{{1}}(3,1,2){{m_{b}}}^{3}}{Z}}+40\,{\frac{\bar{I}%
_{4}^{[0,1]}(3,2,2){{m_{b}}}^{3}}{Z}}+40\,{\frac{\bar{I}_{3}^{[0,1]}(3,2,2){{%
m_{b}}}^{3}}{Z}} \\
&&+20\,{\frac{\bar{I}_{{0}}(2,3,1){{m_{b}}}^{3}}{Z}}-60\,{\frac{\bar{I}_{{3}%
}(4,1,1){{m_{b}}}^{3}}{Z}}-20\,{\frac{\bar{I}_{{3}}(3,1,2){{m_{b}}}^{3}}{Z}}
\\
&&-60\,{\frac{\bar{I}_{{4}}(4,1,1){{m_{b}}}^{3}}{Z}}+10\,{\frac{\bar{I}%
_{1}^{[0,1]}(3,2,2){{m_{b}}}^{3}}{Z}}-40\,{\frac{\bar{I}_{{4}}(3,2,1){{m_{b}}%
}^{3}}{Z}} \\
&&-20\,{\frac{\bar{I}_{{4}}(2,2,2){{m_{b}}}^{3}}{Z}}-15\,{\frac{\bar{I}_{{1}%
}(4,1,1){{m_{b}}}^{3}}{Z}}+20\,{\frac{\bar{I}_{{2}}(2,3,1){{m_{b}}}^{3}}{Z}}
\\
&&+5\,{\frac{\bar{I}_{{1}}(3,2,2){{m_{b}}}^{5}}{Z}}-40\,{\frac{\bar{I}_{{3}%
}(3,2,1){{m_{b}}}^{3}}{Z}}-5\,{\frac{\bar{I}_{{0}}(3,2,1){{m_{b}}}^{3}}{Z}}
\\
&&-15\,{\frac{\bar{I}_{{2}}(3,2,1){{m_{b}}}^{3}}{Z}}+20\,{\frac{\bar{I}_{{4}%
}(3,2,2){{m_{b}}}^{5}}{Z}}+20\,{\frac{\bar{I}_{{3}}(3,2,2){{m_{b}}}^{5}}{Z}}
\\
&&-20\,{\frac{\bar{I}_{{3}}(2,2,2){{m_{b}}}^{3}}{Z}}-20\,{\frac{\bar{I}_{{3}%
}(3,2,2){{m_{b}}}^{5}}{{Z}^{3/2}}}-5\,{\frac{\bar{I}_{{0}}(3,2,2){{m_{b}}}%
^{5}}{{Z}^{3/2}}} \\
&&+20\,{\frac{\bar{I}_{{2}}(3,2,1){{m_{b}}}^{3}}{{Z}^{3/2}}}-20\,{\frac{%
\bar{I}_{{4}}(3,2,2){{m_{b}}}^{5}}{{Z}^{3/2}}}-5\,{\frac{\bar{I}_{{2}}(3,2,2)%
{{m_{b}}}^{5}}{{Z}^{3/2}}} \\
&&+40\,{\frac{\bar{I}_{{4}}(2,2,2){{m_{b}}}^{3}}{{Z}^{3/2}}}+20\,{\frac{%
\bar{I}_{{4}}(3,1,2){{m_{b}}}^{3}}{{Z}^{3/2}}}+25\,{\frac{\bar{I}_{{1}%
}(3,1,2){{m_{b}}}^{3}}{{Z}^{3/2}}} \\
&&+15\,{\frac{\bar{I}_{{0}}(4,1,1){{m_{b}}}^{3}}{{Z}^{3/2}}}-40\,{\frac{%
\bar{I}_{4}^{[0,1]}(3,2,2){{m_{b}}}^{3}}{{Z}^{3/2}}}+65\,{\frac{\bar{I}_{{1}%
}(3,2,1){{m_{b}}}^{3}}{{Z}^{3/2}}} \\
&&+5\,{\frac{\bar{I}_{{2}}(3,1,2){{m_{b}}}^{3}}{{Z}^{3/2}}}-40\,{\frac{%
\bar{I}_{1}^{[0,1]}(3,2,2){{m_{b}}}^{3}}{{Z}^{3/2}}}+25\,{\frac{\bar{I}_{{0}%
}(3,2,1){{m_{b}}}^{3}}{{Z}^{3/2}}} \\
&&+60\,{\frac{\bar{I}_{{1}}(4,1,1){{m_{b}}}^{3}}{{Z}^{3/2}}}+60\,{\frac{%
\bar{I}_{{3}}(4,1,1){{m_{b}}}^{3}}{{Z}^{3/2}}}+10\,{\frac{\bar{I}_{{0}%
}(2,2,2){{m_{b}}}^{3}}{{Z}^{3/2}}} \\
&&+60\,{\frac{\bar{I}_{{4}}(4,1,1){{m_{b}}}^{3}}{{Z}^{3/2}}}+10\,{\frac{%
\bar{I}_{{2}}(2,2,2){{m_{b}}}^{3}}{{Z}^{3/2}}}+5\,{\frac{\bar{I}_{{0}}(3,1,2)%
{{m_{b}}}^{3}}{{Z}^{3/2}}} \\
&&+20\,{\frac{\bar{I}_{{3}}(3,1,2){{m_{b}}}^{3}}{{Z}^{3/2}}}+40\,{\frac{%
\bar{I}_{{1}}(2,2,2){{m_{b}}}^{3}}{{Z}^{3/2}}}+40\,{\frac{\bar{I}_{{3}%
}(3,2,1){{m_{b}}}^{3}}{{Z}^{3/2}}} \\
&&-10\,{\frac{\bar{I}_{0}^{[0,1]}(3,2,2){{m_{b}}}^{3}}{{Z}^{3/2}}}-40\,{%
\frac{\bar{I}_{3}^{[0,1]}(3,2,2){{m_{b}}}^{3}}{{Z}^{3/2}}}-10\,{\frac{\bar{I}%
_{2}^{[0,1]}(3,2,2){{m_{b}}}^{3}}{{Z}^{3/2}}} \\
&&+40\,{\frac{\bar{I}_{{3}}(2,2,2){{m_{b}}}^{3}}{{Z}^{3/2}}}+40\,{\frac{%
\bar{I}_{{4}}(3,2,1){{m_{b}}}^{3}}{{Z}^{3/2}}}-20\,{\frac{\bar{I}_{{1}%
}(3,2,2){{m_{b}}}^{5}}{{Z}^{3/2}}} \\
&&+15\,{\frac{\bar{I}_{{2}}(4,1,1){{m_{b}}}^{3}}{{Z}^{3/2}}}-20\,{\frac{%
\bar{I}_{{4}}(3,2,2){{m_{b}}}^{5}}{{Z}^{2}}}-5\,{\frac{\bar{I}_{{1}}(3,2,2){{%
m_{b}}}^{5}}{{Z}^{2}}} \\
&&-20\,{\frac{\bar{I}_{{3}}(3,2,2){{m_{b}}}^{5}}{{Z}^{2}}}+20\,{\frac{\bar{I}%
_{{3}}(3,2,2){{m_{b}}}^{5}}{{Z}^{5/2}}}+20\,{\frac{\bar{I}_{{4}}(3,2,2){{%
m_{b}}}^{5}}{{Z}^{5/2}}} \\
&&+20\,{\frac{\bar{I}_{{1}}(3,2,2){{m_{b}}}^{5}}{{Z}^{5/2}}}+5\,{\frac{%
\bar{I}_{{0}}(3,2,2){{m_{b}}}^{5}}{{Z}^{5/2}}}+5\,{\frac{\bar{I}_{{2}}(3,2,2)%
{{m_{b}}}^{5}}{{Z}^{5/2}}} \\
&&-10\,\bar{I}_{{1}}(2,3,1){{m_{b}}}^{3}+120\,\bar{I}_{{4}}(1,4,1){{m_{b}}}%
^{3}+120\,\bar{I}_{{3}}(1,4,1){{m_{b}}}^{3} \\
&&-20\,\bar{I}_{{4}}(3,2,1){{m_{b}}}^{3}+30\,\bar{I}_{{1}}(1,4,1){{m_{b}}}%
^{3}-10\,\bar{I}_{{1}}(3,2,1){{m_{b}}}^{3} \\
&&+20\,\bar{I}_{3}^{[0,1]}(3,2,2){{m_{b}}}^{3}-20\,\bar{I}_{{3}}(3,2,1){{%
m_{b}}}^{3}-20\,\bar{I}_{{4}}(2,2,2){{m_{b}}}^{3}
\end{eqnarray*}
\begin{eqnarray*}
C^{HQET}_{2}=&&20\,{\frac{\bar{I}_{4}^{[0,1]}(3,2,2){m_{b}}^{3}}{\sqrt{Z}}}-20\,{\frac{%
\bar{I}_{{4}}(3,2,1){m_{b}}^{3}}{\sqrt{Z}}}-120\,{\frac{\bar{I}_{{3}}(1,4,1){%
m_{b}}^{3}}{\sqrt{Z}}} \\
&&+30\,{\frac{\bar{I}_{{0}}(1,5,1){m_{b}}^{3}}{\sqrt{Z}}}+10\,{\frac{\bar{I}%
_{{1}}(3,2,1){m_{b}}^{3}}{\sqrt{Z}}}+20\,{\frac{\bar{I}_{{3}}(2,2,2){m_{b}}%
^{3}}{\sqrt{Z}}} \\
&&+60\,{\frac{\bar{I}_{{2}}(1,4,1){m_{b}}^{3}}{\sqrt{Z}}}+120\,{\frac{\bar{I}%
_{{4}}(1,4,1){m_{b}}^{3}}{\sqrt{Z}}}+20\,{\frac{\bar{I}_{{3}}(3,2,1){m_{b}}%
^{3}}{\sqrt{Z}}} \\
&&+10\,{\frac{\bar{I}_{2}^{[0,1]}(3,2,2){m_{b}}^{3}}{\sqrt{Z}}}-20\,{\frac{%
\bar{I}_{{2}}(3,2,1){m_{b}}^{3}}{\sqrt{Z}}}+15\,{\frac{\bar{I}_{{0}}(3,2,1){%
m_{b}}^{3}}{\sqrt{Z}}} \\
&&-30\,{\frac{\bar{I}_{{1}}(1,4,1){m_{b}}^{3}}{\sqrt{Z}}}-20\,{\frac{\bar{I}%
_{3}^{[0,1]}(3,2,2){m_{b}}^{3}}{\sqrt{Z}}}-5\,{\frac{\bar{I}%
_{0}^{[0,1]}(3,2,2){m_{b}}^{3}}{\sqrt{Z}}} \\
&&-5\,{\frac{\bar{I}_{1}^{[0,1]}(3,2,2){\ m_{b}}^{3}}{\sqrt{Z}}}-20\,{\frac{%
\bar{I}_{{4}}(2,2,2){\ m_{b}}^{3}}{\sqrt{Z}}}-5\,{\frac{\bar{I}_{{1}}(2,2,2){%
\ m_{b}}^{3}}{Z}} \\
&&-60\,{\frac{\bar{I}_{{3}}(4,1,1){m_{b}}^{3}}{Z}}+20\,{\frac{\bar{I}_{{4}%
}(2,2,2){m_{b}}^{3}}{Z}}-20\,{\frac{\bar{I}_{{3}}(2,2,2){m_{b}}^{3}}{Z}} \\
&&+20\,{\ \frac{\bar{I}_{{4}}(3,1,2){m_{b}}^{3}}{Z}}-5\,{\frac{\bar{I}_{{0}%
}(3,2,1){m_{b}}^{3}}{Z}}+40\,{\frac{\bar{I}_{{4}}(3,2,1){m_{b}}^{3}}{Z}} \\
&&-40\,{\frac{\bar{I}_{{4}}(2,3,1){m_{b}}^{3}}{Z}}+40\,{\frac{\bar{I}%
_{3}^{[0,1]}(3,2,2){m_{b}}^{3}}{Z}}+60\,{\frac{\bar{I}_{{4}}(4,1,1){m_{b}}%
^{3}}{Z}} \\
&&-20\,{\frac{\bar{I}_{{4}}(3,2,2){m_{b}}^{5}}{Z}}-40\,{\frac{\bar{I}_{{2}%
}(2,3,1){m_{b}}^{3}}{Z}}+20\,{\frac{\bar{I}_{{1}}(2,3,1){m_{b}}^{3}}{Z}} \\
&&-15\,{\frac{\bar{I}_{{1}}(3,2,1){{m_{b}}}^{3}}{Z}}-20\,{\frac{\bar{I}_{{3}%
}(3,1,2){m_{b}}^{3}}{Z}}-40\,{\frac{\bar{I}_{{3}}(3,2,1){m_{b}}^{3}}{Z}} \\
&&+30\,{\ \frac{\bar{I}_{{2}}(3,2,1){m_{b}}^{3}}{Z}}+20\,{\frac{\bar{I}_{{0}%
}(2,3,1){m_{b}}^{3}}{Z}}-40\,{\frac{\bar{I}_{4}^{[0,1]}(3,2,2){m_{b}}^{3}}{Z}%
} \\
&&+20\,{\frac{\bar{I}_{{3}}(3,2,2){m_{b}}^{5}}{Z}}+40\,{\frac{\bar{I}_{{3}%
}(2,3,1){m_{b}}^{3}}{Z}}-10\,{\frac{\bar{I}_{0}^{[0,1]}(3,2,2){m_{b}}^{3}}{{Z%
}^{3/2}}} \\
&&+20\,{\ \frac{\bar{I}_{{1}}(3,2,1){m_{b}}^{3}}{{Z}^{3/2}}}+15\,{\ \frac{%
\bar{I}_{{1}}(4,1,1){m_{b}}^{3}}{{Z}^{3/2}}}+20\,{\ \frac{\bar{I}_{{3}%
}(3,1,2){m_{b}}^{3}}{{Z}^{3/2}}} \\
&&+40\,{\ \frac{\bar{I}_{{3}}(2,2,2){m_{b}}^{3}}{{Z}^{3/2}}}-20\,{\ \frac{%
\bar{I}_{{3}}(3,2,2){m_{b}}^{5}}{{Z}^{3/2}}}-10\,{\ \frac{\bar{I}%
_{1}^{[0,1]}(3,2,2){m_{b}}^{3}}{{Z}^{3/2}}} \\
&&-20\,{\frac{\bar{I}_{{4}}(3,1,2){m_{b}}^{3}}{{Z}^{3/2}}}-40\,{\frac{\bar{I}%
_{3}^{[0,1]}(3,2,2){m_{b}}^{3}}{{Z}^{3/2}}}-30\,{\frac{\bar{I}_{{2}}(4,1,1){%
m_{b}}^{3}}{{Z}^{3/2}}} \\
&&+10\,{\frac{\bar{I}_{{0}}(2,2,2){m_{b}}^{3}}{{Z}^{3/2}}}-10\,{\frac{\bar{I}%
_{{2}}(3,1,2){m_{b}}^{3}}{{Z}^{3/2}}}-40\,{\frac{\bar{I}_{{4}}(2,2,2){m_{b}}%
^{3}}{{Z}^{3/2}}} \\
&&+25\,{\frac{\bar{I}_{{0}}(3,2,1){m_{b}}^{3}}{{Z}^{3/2}}}+40\,{\frac{\bar{I}%
_{4}^{[0,1]}(3,2,2){m_{b}}^{3}}{{Z}^{3/2}}}-40\,{\frac{\bar{I}_{{4}}(3,2,1){%
m_{b}}^{3}}{{Z}^{3/2}}} \\
&&-60\,{\frac{\bar{I}_{{4}}(4,1,1){m_{b}}^{3}}{{Z}^{3/2}}}-20\,{\frac{\bar{I}%
_{{2}}(2,2,2){m_{b}}^{3}}{{Z}^{3/2}}}+60\,{\frac{\bar{I}_{{3}}(4,1,1){m_{b}}%
^{3}}{{Z}^{3/2}}} \\
&&+15\,{\frac{\bar{I}_{{0}}(4,1,1){m_{b}}^{3}}{{Z}^{3/2}}}+20\,{\frac{\bar{I}%
_{2}^{[0,1]}(3,2,2){m_{b}}^{3}}{{Z}^{3/2}}}+5\,{\frac{\bar{I}_{{0}}(3,1,2){%
m_{b}}^{3}}{{Z}^{3/2}}} \\
&&+10\,{\frac{\bar{I}_{{1}}(3,1,2){m_{b}}^{3}}{{Z}^{3/2}}}-5\,{\frac{\bar{I}%
_{{1}}(3,2,2){m_{b}}^{5}}{{Z}^{3/2}}}+10\,{\frac{\bar{I}_{{1}}(2,2,2){m_{b}}%
^{3}}{{Z}^{3/2}}} \\
&&-40\,{\frac{\bar{I}_{{2}}(3,2,1){m_{b}}^{3}}{{Z}^{3/2}}}+40\,{\frac{\bar{I}%
_{{3}}(3,2,1){m_{b}}^{3}}{{Z}^{3/2}}}+10\,{\frac{\bar{I}_{{2}}(3,2,2){m_{b}}%
^{5}}{{Z}^{3/2}}} \\
&&-5\,{\frac{\bar{I}_{{0}}(3,2,2){m_{b}}^{5}}{{Z}^{3/2}}}+20\,{\frac{\bar{I}%
_{{4}}(3,2,2){m_{b}}^{5}}{{Z}^{3/2}}}+20\,{\frac{\bar{I}_{{4}}(3,2,2){m_{b}}%
^{5}}{{Z}^{2}}} \\
&&-20\,{\ \frac{\bar{I}_{{3}}(3,2,2){m_{b}}^{5}}{{Z}^{2}}}-10\,{\ \frac{%
\bar{I}_{{2}}(3,2,2){m_{b}}^{5}}{{Z}^{5/2}}}+20\,{\ \frac{\bar{I}_{{3}%
}(3,2,2){m_{b}}^{5}}{{Z}^{5/2}}} \\
&&-20\,{\ \frac{\bar{I}_{{4}}(3,2,2){m_{b}}^{5}}{{Z}^{5/2}}}+5\,{\ \frac{%
\bar{I}_{{0}}(3,2,2){m_{b}}^{5}}{{Z}^{5/2}}}+5\,{\ \frac{\bar{I}_{{1}}(3,2,2)%
{m_{b}}^{5}}{{Z}^{5/2}}} \\
&&+20\,\bar{I}_{3}^{[0,1]}(3,2,2){m_{b}}^{3}+20\,\bar{I}_{{4}}(2,2,2){m_{b}}%
^{3}-40\,\bar{I}_{{3}}(2,3,1){m_{b}}^{3} \\
&&+120\,\bar{I}_{{3}}(1,4,1){m_{b}}^{3}+40\,\bar{I}_{{4}}(2,3,1){m_{b}}%
^{3}-20\,\bar{I}_{{3}}(2,2,2){m_{b}}^{3} \\
&&+20\,\bar{I}_{{4}}(3,2,1){m_{b}}^{3}-20\,\bar{I}_{4}^{[0,1]}(3,2,2){m_{b}}%
^{3}-20\,\bar{I}_{{3}}(3,2,1){m_{b}}^{3}
\end{eqnarray*}
where
\begin{eqnarray}
\bar{I}_n^{[i,j]} (a,b,c) =\frac{(2m_b)^{i+j}}{(\sqrt{z})^{j}}
\left( T_1^2 \right)^i \left( T_2^2 \right)^j
\frac{d^i}{d\left( T_1^2 \right)^i} \frac{d^j}{d\left( T_2^2 \right)^j} %
\left[\left( T_1^2 \right)^i \left( T_2^2 \right)^j \bar{I}_n(a,b,c) \right]%
~.  \nonumber
\end{eqnarray}

\end{document}